\RequirePackage{fix-cm}
\documentclass[smallextended]{svjour3}       
\smartqed  
\usepackage{natbib}
\usepackage{amsmath,amsfonts}
 \usepackage{booktabs}
\usepackage{amssymb}
\usepackage{algorithmic}
\usepackage{graphicx}
\usepackage{textcomp}
\usepackage{multirow}
\usepackage[table,xcdraw]{xcolor}
\usepackage[hidelinks]{hyperref}
\usepackage{adjustbox}
\usepackage{rotating}
\usepackage{listings}
\usepackage{xcolor}
\usepackage{longtable} 
\lstdefinelanguage{json}{
    basicstyle=\ttfamily\small,
    numbers=left,
    numberstyle=\tiny\color{gray},
    stepnumber=1,
    numbersep=8pt,
    showstringspaces=false,
    breaklines=true,
    frame=lines,
    backgroundcolor=\color{lightgray!10},
    literate=
     *{0}{{{\color{blue}0}}}{1}
      {1}{{{\color{blue}1}}}{1}
      {2}{{{\color{blue}2}}}{1}
      {3}{{{\color{blue}3}}}{1}
      {4}{{{\color{blue}4}}}{1}
      {5}{{{\color{blue}5}}}{1}
      {6}{{{\color{blue}6}}}{1}
      {7}{{{\color{blue}7}}}{1}
      {8}{{{\color{blue}8}}}{1}
      {9}{{{\color{blue}9}}}{1}
      {:}{{{\color{red}:}}}{1}
      {,}{{{\color{red},}}}{1}
      {\{}{{{\color{brown}\{}}}{1}
      {\}}{{{\color{brown}\}}}}{1}
      {[}{{{\color{brown}[}}}{1}
      {]}{{{\color{brown}]}}}{1},
}

\colorlet{shadecolor}{gray!10}

\newcommand{\finding}[1]{%
\vspace{-1pt}
\begin{center}
\fcolorbox{lightgray!50}{shadecolor}{\parbox{0.96\textwidth}{\small  
\emoji{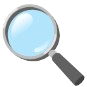}\textcolor{blue}{\findid.} #1}}
\end{center}
\vspace{-2pt}
}
\newcommand*{\finRef}[1]{\textcolor{blue}{\small[{\emoji{mag}}#1]}}

\newcommand{\summary}[1]{%
\begin{center}
\fcolorbox{lightgray}{shadecolor}{\parbox{0.96\textwidth}{\small  \emoji{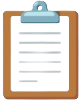} \textbf{Summary:} #1}}
\end{center}
\vspace{-2pt}
}

\newcommand*{\img}[1]{%
    \raisebox{-.3\baselineskip}{%
        \includegraphics[
        height=\baselineskip,
        width=\baselineskip,
        keepaspectratio,
        ]{#1}%
    }%
}
\newcommand*{\bimg}[1]{%
    \raisebox{-.3\baselineskip}{%
        \includegraphics[
        height=\baselineskip,
        width=\linewidth,
        keepaspectratio,
        ]{#1}%
    }%
}
\newcommand*{\emoji}[1]{%
    \img{#1.png}
}
\newcommand*{\citesrf}[1]{\citet{#1}}
\newcommand*{\citeyearp}[1]{(\citeyear{#1})}
\newcommand*{\textpos}[1]{\textcolor{black}{\textbf{#1}}}

\newcommand*{\etals}[1]{
\citesrf{#1}
}
\newcommand*{\pp}[0]{\textit{\%p}}

\newcommand*{\OD}[0]{\textcolor{teal!70}{$\mathcal{D}_{OD}$}}

\newcommand*{\CC}[0]{\textcolor{violet!60}{$\mathcal{D}_{CC}$}}
\newcommand*{\RH}[0]{\textcolor{red!60}{$\mathcal{D}_{RH}$}}

\newcommand*{\crc}[0]{\textit{CR comment}}
\newcommand*{\crcs}[0]{\textit{CR comments}}

\newcommand*{\secref}[0]{Section}

\newcommand*{\BoW}[0]{\textit{Bag-of-Words}}
\newcommand*{\SO}[0]{\textit{SOword2vec}}
\newcommand*{\FA}[0]{\textit{fast}Text}
\newcommand*{\iNewW}[0]{\img{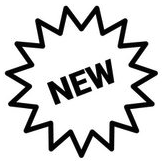}}
\newcommand*{\iNew}[0]{\img{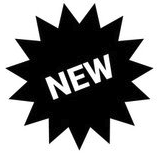}}
\newcommand*{\FALSE}[0]{\bimg{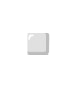}}
\newcommand*{\TRUE}[0]{\textbf{\bimg{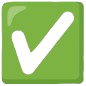}}}

\newcommand*{\iDB}[0]{\img{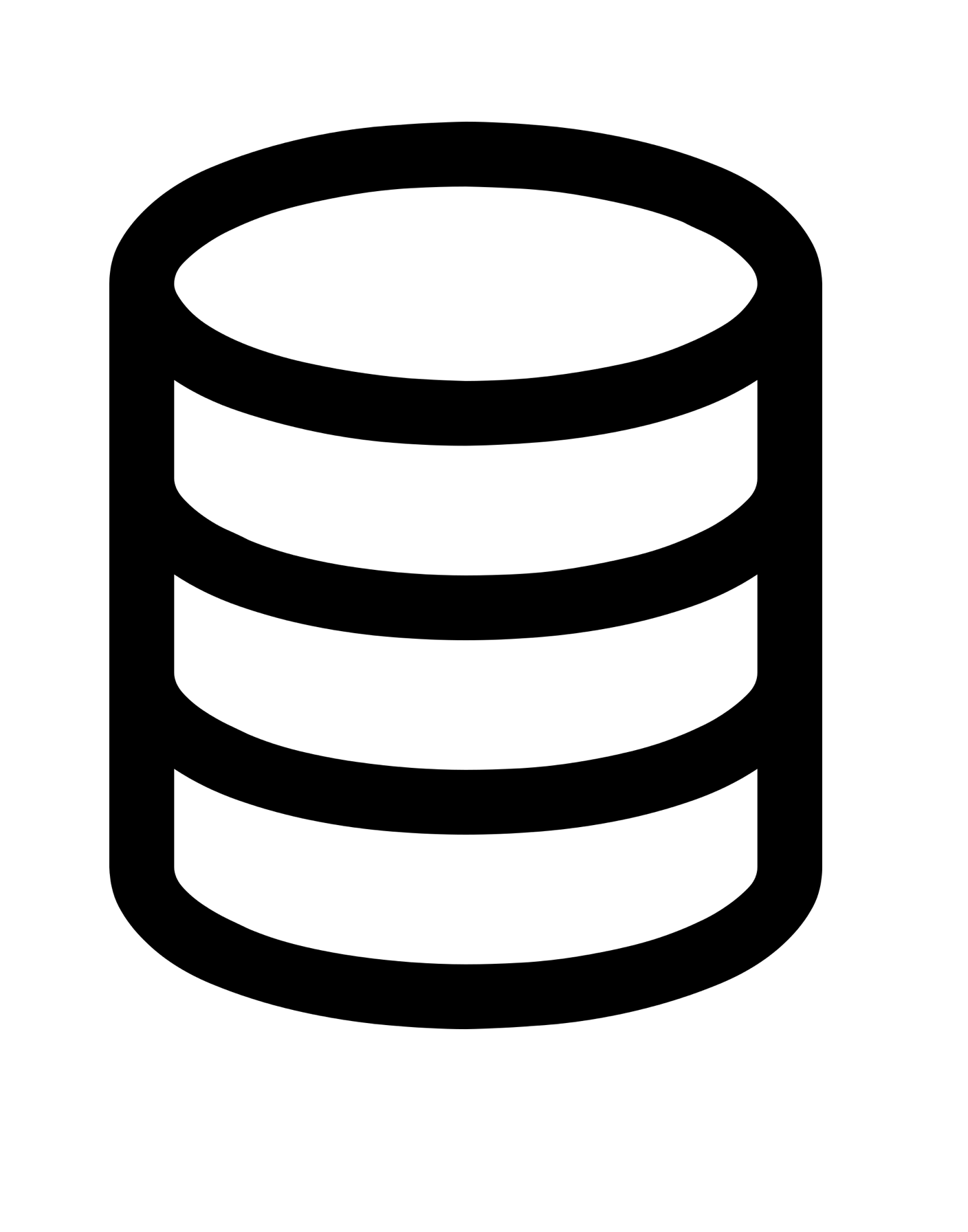}}
\newcommand*{\egcrc}[1]{\textcolor{darkgray}{{\small `\texttt{#1}'}}}
\newcounter{finid}

\newcounter{sbn}
\newcounter{sbsbn}

\newcommand{\findid}{\stepcounter{finid}\thefinid}

\journalname{Empirical Software Engineering}
\begin{document}

\title{Hold On! Is My Feedback Useful? Evaluating the Usefulness of Code Review Comments 
}

\titlerunning{Hold On! Is My Feedback Useful? Evaluating the Usefulness of CR Comments}        

\author{Sharif Ahmed         \and
        Nasir U. Eisty 
}


\institute{Sharif Ahmed \at
              Dept. of CS, Boise State University, Boise, ID, USA \\
              \email{sharifahmed@u.boisestate.edu}           
           \and
           Nasir U. Eisty \at
              Dept. of CS, Boise State University, Boise, ID, USA \\
              \email{nasireisty@boisestate.edu} 
}

\date{
}

\maketitle

\begin{abstract}
\emph{Context:}
In collaborative software development, the peer code review process proves beneficial only when the reviewers provide useful comments.
\emph{Objective:}
This paper investigates the usefulness of Code Review Comments (\crcs) through textual feature-based and featureless approaches.
\emph{Method:}
We select three available datasets from both open-source and commercial projects.
Additionally, we introduce new features from software and non-software domains. 
Moreover, we experiment with the presence of jargon, voice, and codes in \crcs\ and classify the usefulness of \crcs\ through featurization, bag-of-words, and transfer learning techniques.
\emph{Results:}
Our models outperform the baseline by achieving state-of-the-art performance. 
Furthermore, the result demonstrates that the commercial gigantic LLM, GPT-4o, or non-commercial 
naive featureless approach, Bag-of-Word with TF-IDF, is more effective for predicting the usefulness of \crcs. 
\emph{Conclusion:}
The significant improvement in predicting usefulness solely from \crcs\ escalates research on this task. 
Our analyses portray the similarities and differences of domains, projects, datasets, models, and features for predicting the usefulness of \crcs.

\end{abstract}

\begin{figure*}
    \centering
   \centerline{\includegraphics[trim={0 0cm 0 0cm},clip, width=\linewidth]{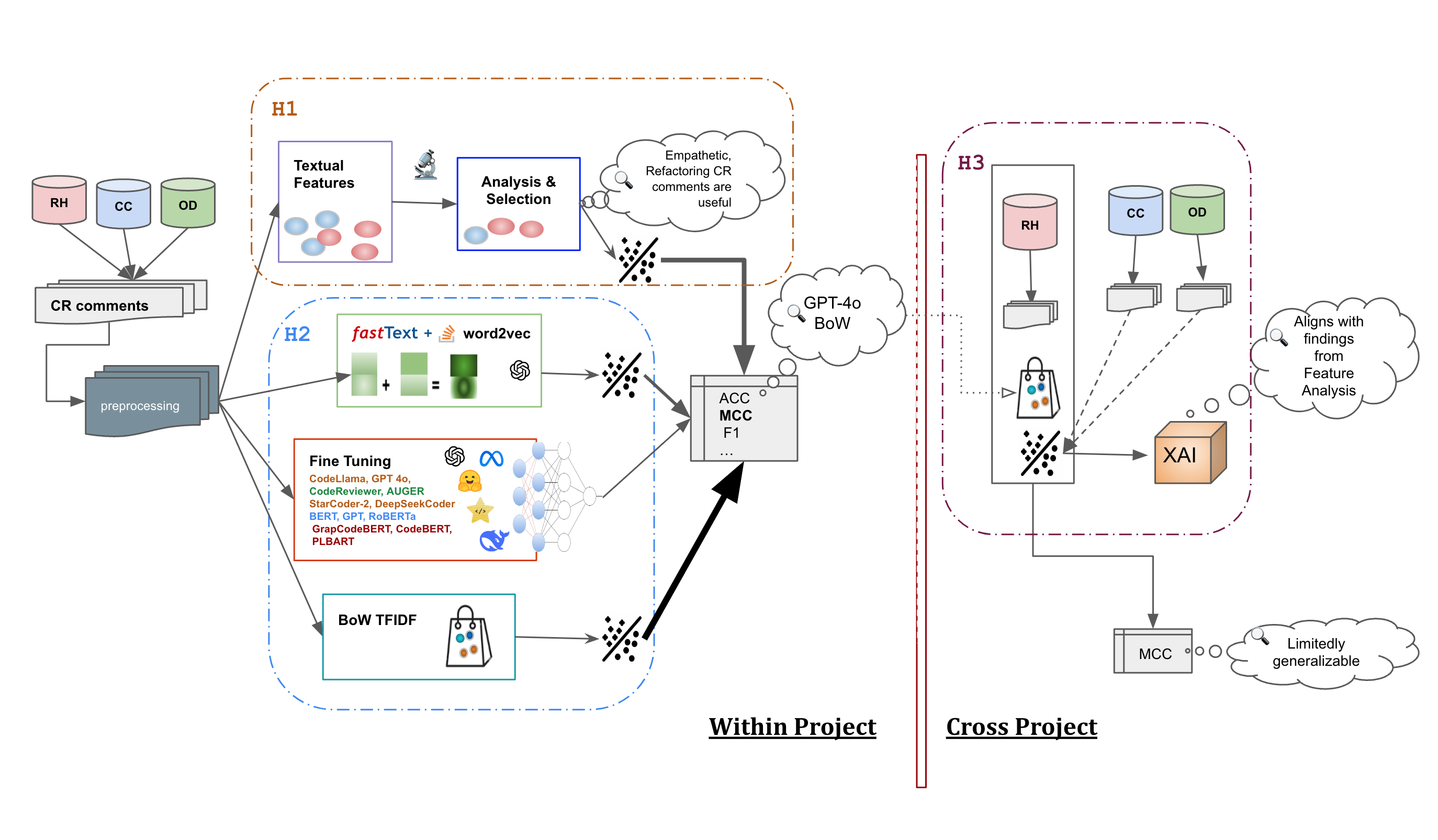}}
    \caption{Overview of our Work}
    \label{fig_approach}
\end{figure*}
\newpage
\section{Introduction}
\label{sec_introduction}

Cohen, known for coining the term \emph{Modern Code Review (MCR)}, emphasizes that authors cannot effectively proofread their own work due to their close proximity to the concepts and the tendency to overlook errors that have become familiar~\citep{cohen2010modern}. 
This self-proofreading problem also applies to coders during the code review process. 
Cohen's research reveals that when code authors review their own work, they only manage to identify approximately 50\% of the defects that an external reviewer would have found.
Even in pre-MCR days, Linus Torvalds told that given enough eyeballs, all bugs are shallow \citep{raymond1999cathedral}.
Furthermore, peer code review is crucial despite its significant time costs. 
According to ~\citesrf{bosu2016process}, developers spend at least a tenth of their time engaging in peer code review activities in both industry and open-source development. 
However, given its essential role, the code review process should be beneficial in enhancing overall team efficiency and project quality.

The core of the MCR process lies in the \crcs\footnote{  ``\crcs'' and ``Code Review Comments'' are interchangeably used in this paper}. 
The effectiveness of the code review process is heavily dependent on the usefulness of these \crcs. 
\citesrf{kononenko2016code} also highlights the significance of useful \crcs\ in an empirical study on the Mozilla code review process.
Despite their importance, empirical findings from \citesrf{bosu2015characteristics} reveal that more than 1 out of 3 \crcs\ at Microsoft are not useful. 
To mitigate this issue, researchers develop models to predict the usefulness of \crcs\ \citep{pangsakulyanont2014assessing,bosu2015characteristics, rahman2017predicting, meyers2018dataset, hasan2021usingCRA}.
These models aiming to predict the usefulness of \crcs\ have taken into account various feature aspects, such as textual (i.e., \crcs\ only), review activities, and developers' experience~\citep{ahmed2023exploring}. 
However, these models may not work well for new developers, new projects, or new companies due to the absence or limited availability of experience records and review activities \citep{ahmed2023exploring}.
Moreover, the association between \crcs' usefulness and non-textual feature (i.e., developers' experience) is conflicting.
Specifically, two studies conducted 
in the past~\citep{bosu2015characteristics,rahman2017predicting} find a positive, whereas a study on OpenDev~\citep{turzo2023makes} finds a negative association. 
Therefore, the only feature aspect that remains consistently available is textual.

After in-depth exploration, we identify three potential avenues from the literature to classify usefulness more effectively, which lead us to introduce new textual features for this task. 
First, 
\citesrf{trenches2018laura} recommends that reviewers offer constructive and respectful \crcs. 
We can capture the style or presentation of \crcs\ with parts-of-speech and other shallow syntactical features~\citep{schneider2016differentiating,zhou2020survey}. 
Second, ~\citesrf{sadowski2018modern} uncovers tone and power dynamics that can result in a breakdown in social interactions or communication between code authors and reviewers in a study at Google. 
We can arrest the tone and power of \crcs\ with models from other domains~\citep{Buechel18emnlp,hosseini2021takes}. 
Third,  during the process of code review and quality improvement, reviewers can also address technical debts and propose refactoring solutions. 
For determining technical suggestions in \crcs\, we can take advantage of the jargon from technical debt, security, refactoring works~\citep{potdar2014exploratory-satd,yu2023security,refactoring2023guru}. 
In summary, we hypothesize that ``\textbf{H1: our proposed textual features that capture language presentation, social interactions, and technical density derived from \crc\ texts effectively classify the usefulness of \crcs.}"
To test this hypothesis, we experiment with new and existing textual features to better understand their role in classifying the usefulness of \crcs.

For text classification, there exist approaches that generate features in their own ways, eliminating the need for laborious feature extraction and selection tasks.
These featureless approaches, including \textit{bag-of-words} and transfer learning techniques, have not been explored yet for predicting the usefulness of \crcs. 
Advanced large pre-trained language models and transformers have the capability to capture both semantic and syntactic relationships. 
Moreover, \citesrf{efstathiou2018code} point out that the semantics of \crcs\ can help predict the usefulness. 
Inspired by this fact, we hunt through the semantics of \crcs\ and put forth the hypothesis that ``\textbf{H2: any feasible cutting-edge featureless technique that considers semantics enhances the measurement of usefulness}."
With this hypothesis in mind, we aim to measure the usefulness of \crcs\ using a range of approaches, from simple \textit{bag-of-words} to more sophisticated domain-agnostic and software domain-specific pre-trained models, as well as popular transformers.

As the code review processes practiced at different teams, organizations, or locations share the same goal to improve the code quality, we hypothesize that ``\textbf{H3: the best-performing usefulness prediction model trained on \crcs\ from a project/ domain is capable of predicting the usefulness of \crcs\ from another project/ domain.}''
Due to the restrictions or unavailability of code and data, the prior works could not verify the generalizability of their models on available related datasets~\citep{ahmed2023exploring}.
But experiments in our study, both handcrafted feature-based (\textbf{H1}) and featureless (\textbf{H2}) approaches, solely take the \crcs-text from available datasets.
This approach allows us to investigate the cross-project generalization of available datasets from different projects. 
To this end, we train the best model on a dataset and evaluate it on other datasets.
To obtain an explanation from our cross-project analysis, we opt for the eXplainable AI (XAI) technique. 
These explanations of the trained models can be considered as interviewing models with data, a proxy for interviewing developers to understand the rationale.

In this paper, we expand upon the existing textual features used to assess the usefulness of \crcs\ by introducing new features for analysis. 
Additionally, we explore featureless techniques for predicting usefulness. 
Subsequently, we conduct evaluations of our prediction models using both old and new evaluation metrics in various settings.
Furthermore, we employ an appropriate \textit{eXplainable} AI tool to elucidate our best models and present our findings from different approaches. 
Our feature analyses provide data-driven perceptions of usefulness from both commercial and open-source projects. 
The overview of our approach is shown in Fig.~\ref{fig_approach}.
Through our feature-based and featureless methods, we surpass baseline performance by \textbf{9\pp-18\pp\ }and 
\textbf{13\pp-42\pp\ }
respectively, as measured by new effective \textit{Matthew's Correlation Coefficient} metric.
Our cross-project evaluation and best model explanation highlight the generalizability of our models trained on different sources. 
Many of our findings from feature analysis and models' explanations support each other.

 The main contributions of this paper are as follows:
\begin{enumerate}
    
    \item New textual features from new and old aspects, their statistical analysis, and feature-based usefulness prediction models.
    \item A thorough featureless approach to predict usefulness from classic\textit{ bag-of-words} all the way to \textit{pre-trained }and\textit{ transformer} language models.
    \item  Evaluation of models within \& cross datasets from diverse sources.
    \item  Explanation of the best models using \textit{eXplainable} AI tool.
    \item  New benchmark scores for usefulness prediction on all datasets.

\end{enumerate}

This paper is structured as follows. Section~\ref{sec_relatedwork} provides a brief overview of \crcs, the notion of their usefulness, predictions of their usefulness, and related work from the literature. 
Section~\ref{sec_methodology} describes our approach to investigate our hypotheses. 
Section~\ref{sec_results} \& \ref{sec_discussion} reports and discusses the results of our investigations. 
Section~\ref{sec_threats} covers the validity threats, and Section~\ref{sec_conclusion} concludes.
Additionally, the artifacts of this paper are included in Section~\ref{sec_availability}.



\newpage
\section{Background \textcolor{black}{and Related Works}}
\label{sec_relatedwork}

This section briefly describes the background, notion, datasets, and prediction of useful Code Review comments (\crcs) from existing literature.

\subsection{Comments in MCR}
The primary objective of Modern Code Review (MCR) is to enhance code quality while saving time and cost notably. 
MCR involves two main stakeholders: the ``author" or ``patch submitter," who requests the review of a code change, and one or more ``peer(s)," ``teammate(s)," ``reviewer(s)," or ``commenter(s)" responsible for reviewing the change.
The MCR process begins with the author uploading a change in the codebase, referred to as a ``patch," ``code-changes," ``diff," or ``change-set." Subsequently, the reviewer(s) are assigned to review the change-set and provide feedback to the author to improve the overall code quality. 
According to \citesrf{davila2021systematic}, review feedback is the most extensively studied internal outcome of the MCR process.
Our study focuses on review feedback provided by a reviewer that is known as Code Review Comments (\crcs), which is completely different than the code comments which are found in the source code of the reviewees or code-authors. 

\subsection{Usefulness of \crcs} 
Our recent study~\citep{ahmed2023exploring} highlights that existing research has defined the usefulness of review feedback or code review comments (\crcs) through various studies~\citep{pangsakulyanont2014assessing, bosu2015characteristics, meyers2018dataset, turzo2023makes}. A few studies mine and annotate \textit{useful/not-useful} \crc\ datasets~\citep{meyers2018dataset, rahman2017predicting}, analyze factors from different aspects~\citep{pangsakulyanont2014assessing, bosu2015characteristics, rahman2017predicting, meyers2018dataset, efstathiou2018code, hasan2021usingCRA}, and employ machine learning classifiers to automatically predict the usefulness of \crcs~\citep{pangsakulyanont2014assessing, bosu2015characteristics, rahman2017predicting, meyers2018dataset, hasan2021usingCRA}.
Most of these works consider findings from a study conducted at Microsoft~\citep{bosu2015characteristics}, suggesting that \crcs\ are useful if they trigger code changes. 
This study takes advantage of the previous usefulness-annotated \crcs\ and textual features derived from \crcs\ to predict the usefulness of \crcs. 

\subsection{Predicting Usefulness of \crcs}
Researchers have built usefulness classifiers for \crcs~\citep{pangsakulyanont2014assessing, bosu2015characteristics, rahman2017predicting, meyers2018dataset, hasan2021usingCRA}.
Their usefulness prediction models have considered three aspects of features, including 
textual or \crc-content~\citep{bosu2015characteristics,hasan2021usingCRA,meyers2018dataset,rahman2017predicting}, developers' experience~\citep{rahman2017predicting,hasan2021usingCRA,bosu2015characteristics}, and review activities~\citep{bosu2016process,hasan2021usingCRA,meyers2018dataset}. 
Interestingly, none of these models could compare their performance with their previous exact models due to the full or partial unavailability of the data and code restricted by Non-Disclosure Agreements~\citep{ahmed2023exploring}. 
However, our study design allows us to compare two prior models that availed their feature values for predicting the usefulness of \crcs\ out of three useful \crcs\ datasets.

\subsection{Experiment Data for Useful \crcs} 
\citesrf{pangsakulyanont2014assessing} gather Gerrit code change commit messages and \crcs\ from the Qt project's \textit{qtbase}. However, their dataset is not publicly available, making it inaccessible for external use.
Similarly, \citesrf{bosu2015characteristics} utilizes five projects from Microsoft.
Unfortunately, \citesrf{bosu2015characteristics} and another study~\citesrf{hasan2021usingCRA} could not make the dataset publicly available due to a Non-Disclosure Agreement (NDA) between the authors and the companies involved.
Despite these limitations, there are three other available datasets.
We use all three available datasets in our experiments.

\iDB\textit{RevHelper (\RH):}
\citesrf{rahman2017predicting} obtains \crcs\ from four commercial subject systems of a company using the GitHub API. 
The authors manually annotate each \crc\ as either useful or not, based on the definition of usefulness provided by \citesrf{bosu2015characteristics}.
However, due to an NDA between the authors and the company, they release a partial dataset that excludes the code changes. 
The dataset contains 879 \textit{useful} and 602 \textit{not-useful} \crcs.

\iDB \textit{ChromiumConversations (\CC):}
\citesrf{meyers2018dataset} generates a dataset from code reviews in the Google Chromium Project. 
By utilizing Rietveld's (code review tool used in Chromium Project)  ``Done" click feature, they automatically identify 2,994 \textit{acted-upon} \crcs, which are considered \textit{useful}. Additionally, the authors manually identify 800 \textit{not-acted-upon} \crcs.
In summary, the dataset comprises a total of 2,994 \textit{acted-upon} (i.e., \textit{useful}) and 800 \textit{not-acted-upon} \crcs.

\iDB \textit{OpenDev (\OD):}
\citesrf{turzo2023makes} obtain a dataset from OpenDev's Nova OSS project using the Gerrit Miner tool. The authors manually annotate the usefulness of \crcs\ based on the code author's explicit appreciation or implicit acknowledgment of the reviewers' comments.
This dataset contains 2,052 \textit{useful} and 602 \textit{not-useful} \crcs.

%


\section{Predicting Usefulness}
\label{sec_methodology}
Here, we describe our approaches to predict the usefulness of code reviewers' \crcs\ written in English from the \crcs\ alone.

\subsection{Data Setup and Pre-Processing}

\subsubsection{Data Setup}
For our experiment, we utilize three established datasets: \RH, \OD, and \CC\ (\secref~\ref{sec_relatedwork}). 
These datasets contain raw \crc\ texts along with usefulness labels. 
We check the papers and artifacts to ensure the datasets consist solely of \crcs\ from reviewers.
During the verification process, we identify 24 duplicate \crcs\ in \RH, 104 in \CC, and 191 in \OD. 
To maintain data integrity and prevent any data leakage during training and testing, we deduplicate the \crcs\ in the datasets.

\noindent\subsubsection{Preprocessing} 
Though \crcs\ are short (median size of 35) in existing datasets, they have words that may not carry meaning to the code authors. 
So, we decide to preprocess the \crcs\ and keep five additional variants of \crcs\ in our experiment: 

\begin{enumerate}
    \item \texttt{comment}: the original \crcs. 
    \item \texttt{code}: snippet and code extracted from \crcs\ (\secref~\ref{new_fts_method})  
    \item \texttt{text}: \crcs\ after removing code, \textit{\#issue, @username, email-id}  
    \item \texttt{text clean}: text after removing URL and markdown links 
    \item \texttt{text tokens}: tokenized text with NLTK's TweetTokenizer 
    \item \texttt{code tokens}: tokenized code with Spiral\footnote{https://github.com/casics/spiral} 

\end{enumerate}

\subsection{Predicting Usefulness with Features}
\phantomsection \label{rh1}

To investigate our first hypothesis, \textbf{H1} (\secref~\ref{sec_introduction}), we introduce a set of new features from both software and non-software domains that could potentially impact developers socially or individually. 
These additional features are combined with existing textual features~\citep{ahmed2023exploring} to create a comprehensive feature-set for classifying useful \crcs. 
It is important to note that, for this investigation, we assume that review context and developer experience are unrelated to our hypotheses, and thus, we exclude them from our consideration.
The complete set of old and new features, along with the selection of salient features and the models used for our classification task, can be found in Table~\ref{tbl_fts_all}. 
We provide detailed descriptions of the old and new features, the methodology for feature selection, and the models utilized for classification in the following subsections.


\newcounter{ft}
\setcounter{ft}{0}
\newcommand{\fid}{\stepcounter{ft}\theft}

\begin{table*}[hbtp]
\renewcommand{\arraystretch}{0.9}
\selectfont
\centering
\caption{Textual Features Used for Classifying Useful \crcs}
\label{tbl_fts_all}
{\tiny
 Here FID prefixes are \textit{V: voice, C: code, T: text,} J\textit{: jargon ;
Citation following Feature names indicates existing features \& \iNew indicates our proposed features for useful \crcs\ prediction.}
.

}
 \resizebox{\textwidth}{!}{%
\begin{tabular}{|
l
|p{0.28\linewidth}
|p{0.71\linewidth}|}  
\hline
ID& Feature Name &  Description \\
\hline

 T\fid &         { avg-words\iNew}      &      Average words per sentence~( Zohu, \citeyear{zhou2020survey}) \\ 
T\fid &         { num-verb\iNew}      &      Counts verbs ~( Zohu, \citeyear{zhou2020survey})\\ 
T\fid &         { avg-stopwords\iNew}      &      Average stop words per sentence~( Zohu, \citeyear{zhou2020survey})\\ 

T\fid &         { word-count \citeyearp{hasan2021usingCRA}}     &      Number of words in a comment.~( Zohu, \citeyear{zhou2020survey})\\ 
T\fid &         { num-chars\iNew}      &      Counts characters in \crc~( Zohu, \citeyear{zhou2020survey})\\ 
T\fid &         { num-determinants\iNew}      &      Counts determinants~( Zohu, \citeyear{zhou2020survey})\\ 
T\fid &         { num-nouns\iNew}      &      Counts nouns ~( Zohu, \citeyear{zhou2020survey})\\ 
T\fid &         { num-adj\iNew}      &      Counts adjectives~( Zohu, \citeyear{zhou2020survey})\\ 
T\fid &         { num-adverb\iNew}      &      Counts adverb~( Zohu, \citeyear{zhou2020survey})\\ 
C\fid &         { prog-words \citeyearp{bosu2015characteristics}}     &      Counts programming words (keywords , identifiers)\citeyearp{hasan2021usingCRA}\\ 
V\fid &         { num-tentative\iNew}      &      Counts the uncertainty words (i.e.\textit{probable, probably, possible, possibly, perhaps, like})~( Zohu, \citeyear{zhou2020survey})\\ 
T\fid &         { stop-word-ratio \citeyearp{rahman2017predicting}}     &      The ratio of stop words in the comment \citeyearp{hasan2021usingCRA}\\ 
T\fid &         { num-sent\iNew}      &      Counts sentences in \crc~( Zohu, \citeyear{zhou2020survey})\\ 
T\fid &         { avg-punct\iNew}      &      Average punctuations per sentence~( Zohu, \citeyear{zhou2020survey})\\ 
C\fid &         { has-out-snippet\iNew}      &      True if the comment has variables and function signatures without any back ticks\\ 
T\fid &         { subjectivity\iNew}      &      0.0 is very objective and 1.0 is very subjective\citeyearp{textblob}\\ 
T\fid &         { avg-chars\iNew}      &      Average characters per word~( Zohu, \citeyear{zhou2020survey})\\ 
T\fid &         { rd-text \citeyearp{bosu2015characteristics,rahman2017predicting}}     &      Flesh Kincaid’s readability measure\\ 
T\fid &         { num-Qmark\iNew}      &      presence of question mark symbol (?) in \crcs~( Zohu, \citeyear{zhou2020survey})\\ 
C\fid &         { code-word-ratio \citeyearp{rahman2017predicting}}     &      The ratio of source code tokens in the comment\citeyearp{hasan2021usingCRA}\\ 
T\fid &         { question-ratio \citeyearp{rahman2017predicting}}     &      Ratio of interrogative sentences in a comment\\ 
V\fid &         { is-confirmatory \citeyearp{hasan2021usingCRA}}     &      If the author responds with ``Done'', ``Fixed'', ``Removed''\\ 
T\fid &         { num-exclamation\iNew}      &      Counts ``!" signs~( Zohu, \citeyear{zhou2020survey})\\ 
T\fid &         { polarity\iNew}      &      the +1 means a very positive statement and 1 means a very negative statement \citeyearp{textblob}\\ 
T\fid &         { num-interjections\iNew}      &      Counts interjections~( Zohu, \citeyear{zhou2020survey})\\ 
T\fid &         { num-propernouns\iNew}      &      Counts proper nouns~( Zohu, \citeyear{zhou2020survey})\\ 
V\fid &         { cr-senti \citeyearp{hasan2021usingCRA}}     &      The sentiment of the comment using SentiCR \citeyearp{ahmed2017senticr}\\ 
V\fid &         { is-toxic\iNew}      &      \etals{sarkar2023toxicr} toxicity analyzer\\ 
V\fid &         { tone\iNew}      &      Provides tone i.e. empathy directions $\in$ \{ Seeking, Providing, No empathy \} ~\citeyearp{hosseini2021takes}\\ 
T\fid &         { yngve \citeyearp{meyers2018dataset}}     &      The Yngve text complexity score for \crc \\ 
T\fid &         { cdensity \citeyearp{meyers2018dataset}}     &      The Content Density score for \crc \\ 
T\fid &         { pdensity \citeyearp{meyers2018dataset}}     &      The Propositional Density score for \crc \\ 
T\fid &         { frazier \citeyearp{meyers2018dataset}}     &      The Frazier text complexity score for \crc \\ 
V\fid &         { informativeness\iNew}      &      The Informativeness score for sentences in the code review comment~\citeyearp{lahiri2015squinky} \\ 
V\fid &         { formality \citeyearp{meyers2018dataset}}      &      The formality score for review comments \\ 
V\fid &         { politeness \citeyearp{meyers2018dataset}}     &      Politeness score for \crc \\ 
V\fid &         { implicature\iNew}      &      The Implicature score for \crc~\citeyearp{lahiri2015squinky} \\ 
V\fid &         { gratitude \citeyearp{hasan2021usingCRA}}      &      If the code author responds with ``Thank you”/``Thanks”\\ 
V\fid &         { distress (power)\iNew}      &      Provides scores based on how worried, upset, troubled, perturbed, grieved, disturbed, alarmed, and distressed a message is\citeyearp{Buechel18emnlp}\\ 
V\fid &         { empathy (power)\iNew}      &      It provides a score based on how warm, tender, sympathetic, softhearted, moved, and compassionate a message is \citeyearp{Buechel18emnlp} \\ 
J\fid &         { density-refact-solu\iNew \citeyearp{refactoring2023guru}}      &      To determine if the reviewer shares any refactoring suggestion \\ 
J\fid &         { density-secdev\iNew  \citeyearp{yu2023security}}      &      To determine if the reviewer addresses any security issues \\ 
J\fid &         { density-refact-prob\iNew  \citeyearp{refactoring2023guru}}      &      To determine if the reviewer addresses any refactoring issues \\ 
J\fid &         { kw-msoft-nu \citeyearp{bosu2015characteristics}}      &      not useful word counts from Microsoft Study \citeyearp{bosu2015characteristics}\\ 
J\fid &         { density-satd\iNew}      &      \citeyearp{potdar2014exploratory-satd} To check if the reviewer addresses any technical debts \\ 
J\fid &         { kw-msoft-u \citeyearp{bosu2015characteristics}}      &      useful word counts from Microsoft Study \citeyearp{bosu2015characteristics} \\ 
J\fid &         { density-refact-xerx\iNew}      &      \citeyearp{alomar2021refactoring-xerox} To see if the reviewer addresses any refactoring issues \\ 
C\fid &         { has-snippet\iNew}      &        True if the \crc\ has code elements inside back-tick quotations \\ 
 
 \hline 
 
 \end{tabular}
 }
 \end{table*}

\subsubsection{Existing Features} 
\label{sec_sub_oldfts}
 
We initially incorporate existing textual features found in our related research. 
Specifically, we consider the features (FID 4,10,12,18,20-22,27,38 in Table~\ref{tbl_fts_all}) from \citesrf{hasan2021usingCRA}, which utilize numerous features from their previous studies~\citep{bosu2015characteristics, rahman2017predicting} and keyword based features FID 44 \& 46 from~\citesrf{bosu2015characteristics}.

Subsequently, we attempt to integrate linguistic features (FID 30-33,35,36 in Table~\ref{tbl_fts_all}) from \citesrf{meyers2018dataset}. 
However, we encounter deprecated dependency models for the \textit{politeness} and \textit{formality} features. 
To address this, we reimplement these features based on the original work by \citesrf{lahiri2015squinky}.
During the process, we identify that certain features such as \textit{cdensity, pdensity, yngve,} and \textit{frazier} require a parser, which is not specified in their paper or artifact~\citep{meyers2018dataset}. 
Consequently, we adopt feature implementations that are adapted with the BLLIP parser\footnote{https://www.nltk.org/\_modules/nltk/parse/bllip.html} to fulfill the requirements and facilitate the integration of these features.

\subsubsection{New Features}
\label{new_fts_method}

In our recent study\citep{ahmed2023exploring}, we categorized \crcs\ into two aspects: \textit{code} and \textit{text}. 
Building upon this categorization, we conduct a thorough examination of the raw \crcs\ and recognize the existence of two additional aspects: \textit{jargon} and \textit{voice}. 
This newly identified aspect offers valuable information, prompting us to derive new features based on each aspect in the following manner.
We describe our proposed features within \iNewW~existing and \iNew~new aspects.
With our proposed features in addition to existing ones, we experiment the robustness of feature-based classification of useful \crcs.

\paragraph{\iNewW Code} 

Apart from existing code-related textual features, we introduce two new features that contribute to describing and classifying the usefulness of \crcs.
The majority of reviewers quote code portions enclosed with either 1 or 3 back-ticks, denoted as \textit{\`{}code\_snippet\`{}} or \textit{ \`{}\`{}\`{}code\_snippet\`{}\`{}\`{}}). 
However, we observe that there are also instances of different code syntax, variables, and functions found outside the back-ticks. 
 To capture these code snippets, we utilize the regular expression: \textsc{\small (\`{}~\{3\}(.*?)\`{}~\{3\})|(\`{}~\{1\}(.*?)\`{}~\{1\})}%
.

For \textit{camelCase} and \textit{snake\_case} identifiers we use: \textbf{\scriptsize \texttt{[a-z\_\$0-9A-Z]+[A-Z]+[a-z\_0-9]+'}}
. 

And for \textit{function signatures} with parameters we use: \textbf{\scriptsize $[a-zA-Z][a-zA-Z0-9\_.]*\backslash([a-zA-Z0-9\_, ]*\backslash)$}.

Furthermore, the back-ticks (\`{}) serve as visual indicators to readers that a particular segment is a chunk of code, enhancing the appearance and readability~\citep{healy2018data}. 
As a result, we introduce two features based on this information: \textbf{has out-snippets} and \textbf{has snippets} (FID 15 and 48 in Table~\ref{tbl_fts_all}). 
These features capture the presence of code snippets within and outside the back-ticks, respectively.

\paragraph{\iNew Voice} 

Upon extracting the code elements, the review comments transform into natural text. 
The natural text does not have the voice. 
However, the text content can express the voice through how the content is delivered. 
For example, a simple interrogative question may have mild or harsh voices. 
Similarly, some may put multiple question marks or multiple exclamations on their questions.
Moreover, subjective text reflects the speaker’s personal voice, opinions, emotions, and perspective, while the objective text is more tied to facts, information, or observations. 
So, we plan to measure these factors leveraging 
existing natural language processing research~\citep{zhou2020survey,hosseini2021takes,Buechel18emnlp}. 

Unlike previous works that treat \crc\ sentiments as a single entity, combining subjectivity and emotions, we take a different approach. 
To evaluate the subjectivity of comments separately, we incorporate TextBlob\footnote{https://pypi.python.org/pypi/TextBlob}'s \textbf{subjectivity} scores in addition to the existing \textbf{polarity} scores.
Additionally, we recognize the significance of assessing the \textbf{toxicity} of \crcs\ and, for this purpose, utilize the \textit{ToxiCR} model~\citep{sarkar2023toxicr} in addition to existing state-of-the-art SE-emotion features mentioned in Section~\ref{sec_sub_oldfts}.

Drawing insights from Sadowski's empirical study at Google~\citep{sadowski2018modern}, which highlights how the `tone' and `power' of \crcs\ could impact the review process, we integrate the \textbf{tone} \textit{(empathy direction: providing empathy, seeking empathy, or none)} and \textbf{power} \textit{(empathy score and distress scores)} features. 
We adopt the \textit{Empathy Direction} model~\citep{hosseini2021takes} and the \textit{Empathetic Reaction} models~\citep{Buechel18emnlp} for this purpose.
Furthermore, we decide to include two additional linguistic features, \textit{implicature} and \textit{informativeness} (FID 34 \& 37), from the already adopted model for existing features FID 35 \& 36.

\paragraph{\iNewW Text} 

\citesrf{trenches2018laura} suggest that \crcs\ need to be constructive and respectful during code review practice. 
We can capture the style or presentation of \crcs\ with parts-of-speech and other shallow syntactical features~\citep{schneider2016differentiating,zhou2020survey}.

To capture the \crcs\ style, we consider syntactic features from \citesrf{zhou2020survey}, such as Parts-of-Speech, words, sentences, etc. (FID 1,3,5-9,13,14,17,22,23,25,26 in Table~\ref{tbl_fts_all}). 
To refer to all these features collectively, we used the term \textit{syntactic features} in our experiment.

\paragraph{\iNew Jargon}
Code review process involves technical discussions through \crcs.
We can identify these technical jargon from software engineering research that have shared lists of words in the context of technical debt, security, and refactoring, which also fulfill code review goals.

According to information shared by Microsoft developers in \citesrf{bosu2015characteristics}, a team is dedicated to manually monitoring issues in \crcs\ every week. 
Inspired by this practice, we incorporate several keyword-based features 
that identify and tally the jargonic words 
to aid in classifying useful \crcs.
Initially, 
we introduce the \textbf{(SA)TD density} feature by leveraging \textit{Self-Admitted Technical Debt (SATD)} keywords from \citesrf{potdar2014exploratory-satd} to assess whether the reviewer's \crcs\ pertain to technical debts.
For example, a reviewer has commented the following \crc.
\egcrc{ just abandon it. }
Here, it has the `abandon' (SA)TD content keyword once, which makes SATD density score 1.
Furthermore, we introduce the \textbf{refactoring density} feature to determine if the reviewer addresses any refactoring issues. 
To achieve this, we utilize refactoring keywords from an industrial study at Xerox by \citesrf{alomar2021refactoring-xerox}. 

Additionally, we obtain the \textbf{security density} feature from the work of \etals{yu2023security} to identify if the reviewer discusses security defects in their comments.
In addition to the above features, we derive two more refactoring features from an educational website~\citep{refactoring2023guru}. 
We find concise definitions of 
19 refactoring problems and their corresponding solutions on this website. 
We extract jargon from these definitions and develop two features, namely \textbf{refactoring problem density} and \textbf{refactoring solution density}.

Finally, we examine the \textit{unigram,} the \textit{unigram$_{ \notin stopwords}$,} the \textit{lemmatized-unigram$_{ \notin stopwords}$}, and \textit{bigram} (for sources with two or more words~\citep{potdar2014exploratory-satd, refactoring2023guru}) and check their presence to calculate the respective densities of these keywords. 
These jargon-density-based features collectively contribute to the classification of useful \crcs.

\subsubsection{Feature Analysis} 

After extracting both existing and proposed features from Section~\ref{sec_sub_oldfts}~\&~\ref{new_fts_method}, we analyze all the features, as listed in Table~\ref{tbl_fts_all}, using the \textit{Pearson Correlation}. 
This analysis allows us to gain insights into the relationships between all features and the usefulness of \crcs\ on the chosen datasets.

\paragraph{Within-project}

We perform a comparative analysis of input features with respect to the usefulness of \crcs, similar to the approach taken in \citesrf{rahman2017predicting}. 
To determine the differences between features associated with useful and non-useful \crcs, we execute the \textit{Mann Whitney Wilcoxon test}. 
Additionally, we use \textit{Cohen's D} as an effect-size measure to quantify the magnitude of these differences.

\paragraph{Cross-Project}
\label{fts_analysis_cross_proj}
To statistically evaluate the distinguishing power of features in identifying the usefulness of \crcs\ from Commercial~\citep{rahman2017predicting} and Open Source~\citep{meyers2018dataset, turzo2023makes} environments, we perform the 
statistical t-tests between the pairs of all the datasets.
\begin{itemize}
    \item First, we take features' correlations to the usefulness label from two datasets.
    Here we consider \textit{Pearson Correlation} measure.
    \item Second, we pass two sets of 48 feature-correlations (Table~\ref{tbl_fts_all}) from two of the datasets to \textit{Wilcoxon signed-rank test}~\citep{wilcoxon1970critical}, non-parametric version of the paired T-test.
    \item Third, we quantify their effect size using \textit{Cohen's D}.
    \item Fourth, we repeat the above three steps between \CC-\RH, \RH-\OD, and \OD-\CC.
\end{itemize}

This test helps determine if the features exhibit significant differences between the two types of environments.

\subsubsection{Feature Selection} We select three feature selection approaches to classify usefulness using hand-crafted textual features: 

\paragraph{Significant Features}

We perform a comparative analysis of input features with respect to the usefulness of \crcs, similar to the approach taken in \citesrf{rahman2017predicting}. 
To determine the differences between features associated with useful and non-useful \crcs, we execute the \textit{Mann Whitney Wilcoxon test}~\citep{wilcoxon1970critical}. 
Additionally, we use \textit{Cohen's D}~\citep{cohen2013statistical} as an effect-size measure to quantify the magnitude of these differences. 
After obtaining \textit{p-values} from the above within-project feature analysis, we select the features as significant if they have \textit{p-value<0.05}.

\paragraph{Relevant Features} 
Initially, we conduct feature selection by eliminating highly correlated features (>0.9) among themselves, retaining only the one that exhibited a stronger correlation with the usefulness of \crcs, following a similar approach as described in \citesrf{hasan2021usingCRA}.
Subsequently, we determine the most effective textual features by subjecting them to an analysis against the usefulness of \crcs\ through \textit{Recursive Feature Elimination by Cross Validation (RFECV)} \citep{meyers2018dataset, hasan2021usingCRA}.

\paragraph{Important features}

We calculate the XGBoost feature importance scores based on the computed feature values to select important features. 
Unlike p-values, effect size, and correlation measures, the importance scores lack interpretable threshold values. 
Therefore, we opt to use the 75\% quartile of the obtained scores as the importance threshold.

Once we obtain the features selected from the three approaches and three datasets, we obtain feature sets by observing their appearances in one, two, or all three of the datasets
. 
Additionally, we include all the features to train the classifiers, ensuring a comprehensive exploration of feature combinations.

\subsubsection{Classification}
\phantomsection \label{rh1clf}

Using the selected features, we perform stratified 10-fold cross-validation on our datasets using three different classifiers: Random Forest (RF)~\citep{rahman2017predicting}, Gaussian Naive Bayes (NB), and Logistic Regression (LR)~\citep{meyers2018dataset}.
We keep the same hyperparameters of the baseline classifiers for experimenting with our models.

\subsection{Predicting Usefulness without Features}
\phantomsection \label{rh2}

To test our second hypothesis, \textbf{H2} (\secref~\ref{sec_introduction}), we explore a range of classic to very complex featureless Natural Language Processing (NLP) models.
Initially, we employ simple featureless techniques based on \textit{Bag of Words} to classify the usefulness of \crcs. 
Subsequently, we utilize the modern featureless approach called ``Transfer Learning," which allows us to leverage pre-existing knowledge from one model and apply it to another without starting from scratch. 
Transfer learning can be achieved through two main approaches: learning features from pre-trained models (e.g., word2vec, Glove, fastText, etc.) and fine-tuning pre-trained models (e.g., BERT, GPT-4, etc.).
The learning of word representations from pre-trained models yields dense vectors for each word, typically ranging from 50 to 1000 dimensions. 
These vectors inherently encapsulate semantic meanings that the models learned from the vast corpora on which they were trained.

\subsubsection{Bag of Words} 
\phantomsection \label{methode_bow}
In this approach, we transform the words from \crcs\ into a vector representation using the concept of a Bag of Word~\citep{feldman2007text}. 
To enhance the significance of distinctive words and reduce the impact of common words, we apply TF-IDF (Term-frequency \& Inverse-document-frequency) as the weighting scheme.
TF-IDF assigns higher weights to words that are frequent in a specific document but rare across the entire dataset.
To further refine the preprocessing, we consider \textit{stopwords} removal\textit{, stemming, }and\textit{ lemmatization }as hyperparameters. 
For \textit{stopwords}, we select both English \textit{stopwords} from libraries like NLTK and Sci-kit, as well as programming-specific \textit{stopwords} from \citesrf{hasan2021usingCRA}.
This process allows us to remove common and domain-specific words that do not carry significant meaning for the classification task.
The complete list of stopwords is shown in Table~\ref{tbl_stopwords}. 
Once the vector representations for \crcs\ have been obtained, we use the classifiers mentioned in Section~\ref{rh1clf} to predict the usefulness of \crcs.

\subsubsection{Embedding from Pre-trained Models} 
\phantomsection \label{methode2v}
This section provides a concise overview of the pre-trained models we select to learn word representation for our \crcs. 
These pre-trained embeddings are heavier than BoW and lighter than medium or large language models regarding computational and memory resource usage.
Their inclusions cover experimenting with all-sized featureless models and techniques.

\paragraph{SE Embedding}  
To represent the text element in our \crcs, we opt for a domain-specific pre-trained model called \SO, which is built on Stack Overflow posts~\citep{efstathiou2018word}. 
We select the \SO\ model for several reasons. 
The communication patterns observed between questioners and answerers on Stack Overflow resemble those between code authors and code reviewers during code reviews. 
Both forms of communication aim to solve problems and enhance code quality, making the \SO\ model's training context highly relevant to the code review platform.
Moreover, using pre-trained models from non-software domains might lead to misleading semantics in the software engineering context~\citep{jongeling2017negative}. 
For instance, Efstathiou et al.~\citep{efstathiou2018word} discover that the word ``cookie" in the software domain is associated with terms like ``sessionid and session." 
In contrast, in the Google News domain, it is related to ``cupcake and oatmeal cookie." 
To minimize such domain noise in our results, we choose the \SO\ pre-trained model, ensuring better alignment with the software engineering context.

\paragraph{{NL} Embedding}
For the code element, we opt for a 
general domain 
pre-trained model called \FA~\citep{fasttext}. 
Additionally, it is beneficial for code snippets where unknown or out-of-vocabulary tokens can be common, compared to our SE-based \SO\ embedding.
The \FA\ utilizes a sub-word model, considering each word as a bag of character n-grams. 
Though it has a shallow contextual understanding than medium or large language models,  it is extremely fast at providing word representations. 

Given that code snippets can often contain unique or rarely occurring identifiers or symbols, \FA's capability to handle unknown tokens makes it a suitable choice for representing the code element in our analysis.

\paragraph{{Commercial} Embedding} Recent successes achieved in many problem-solving cases by commercial large language models (LLMs) and their applications, such as ChatGPT, have motivated us to opt for commercial embeddings. 
For our usefulness classification task, we follow OpenAI's guidelines and choose their recently released paid \textit{Text-embedding-3-small} embedding model to obtain the semantics of our \crcs.
This embedding model provides a long vector of size 1536 for a given \crc.

\paragraph{Classification} 
After obtaining embeddings from the pre-trained models and various combinations for text and code parts, we feed these combined embeddings into the same classifiers mentioned in Section~\ref{rh1clf} to predict the usefulness of \crcs.

\subsubsection{Fine-tuning Transformer Language Models}
To verify our second hypothesis, \textbf{H2}, we lastly conduct experiments with Transfer Learning by fine-tuning popular pre-trained transformer-based models. This approach involves adapting these models to suit our datasets and the specific requirements of our usefulness classification task.

\paragraph{{Model Selection}} 

 We choose the following pre-trained %
models which are trained on non-software and software domain corpora.

\begin{itemize}
    \item  First, we consider general state-of-the-art (SOTA) large language models (LLMs): 
    GPT 4-o (OpenAI\footnote{https://openai.com}, Aug 2024),
    GPT-Base (OpenAI, 2024),  
    RoBERTa~\citep{liu2019roberta}, and 
    BERT~\citep{devlin2018bert}.
    These proven models have succeeded in various Natural Language Processing (NLP) tasks such as  text classification, question-answering, text understanding and text completions across many general and software domains.
    As the \crcs\ are based on natural languages used by software developers, these models are potential for classifying the usefulness of \crcs.

    \item Researchers have pre-trained, fine-tuned, and optimized LLMs specifically for programming-related tasks like code generation, completion, and understanding multiple programming languages.  
    These hybrid and code models are supposed to classify the usefulness of \crcs\ better than the general LLMs. 
    So, we choose code LLMs - 
    StarCoder~2~\citep{lozhkov2024starcoder}, 
    DeepSeek-Coder~\citep{guo2024deepseek}, 
    CodeLlama~\citep{roziere2023code}, 
    PLBart~\citep{ahmad2021unified}, 
    GraphCodeBERT~\citep{guo2020graphcodebert}, and CodeBERT~\citep{feng2020codebert}.

    \item Finally, we choose code-review-related SOTA code LLMs
    AUGER~\citep{li2022auger} and
    CodeReviewer~\citep{li2022codereviewer}.
    These models are focused on automating code review related tasks.
    
    \item 
    \noindent These selected models have 125 Million to 125 Billion parameters. 
The commercial GPT-4o has approximately 1+Trillion parameters.
For closed LLMs, such as GPT-Base and GPT-4o, we choose the largest architectures. 
For open LLMs with less than 1 Billion parameters, we chose their largest model architecure.
For models with 1 Billion to 70 Billions parameters, we find their ~7 Billion parameters architectures perform better than their lower or upper architectures. 
So we choose StarCoder2-7b, DeepSeek-Coder-6.7b, and CodeLlama-7b.
Most of these models use Byte-Pair-Encoding (BPE), some use coding adapted BPE. 
    
\end{itemize}

\paragraph{Fine-tune Setup}

We choose the following 
tokenizer and models via hugging face PyTorch transormers\footnote{https://huggingface.co} for fine-tuning:
``roberta-base'' for RoBERTa~\citep{liu2019roberta}, ``bert-base-uncased'' for BERT~\citep{devlin2018bert}, ``microsoft/ graphcodebert-base'' for GraphCodeBERT~\citep{guo2020graphcodebert} and ``microsoft/ codebert-base'' for CodeBERT~\citep{feng2020codebert}, 
``bigcode/starcoder2-7b'' for StarCoder~2~\citep{lozhkov2024starcoder}, 
``deepseek-ai/deepseek-coder-6.7b-instruct'' for DeepSeek-Coder~\citep{guo2024deepseek}, 
``meta-llama/CodeLlama-7b-hf'' for CodeLlama~\citep{roziere2023code}, 
``uclanlp/plbart-base'' for PLBart~\citep{ahmad2021unified}, 
``SEBIS/code\_trans\_t5\_base\_code\_documentation\_generation \_java \_multitask'' for AUGER~\citep{li2022auger} and
``microsoft/codereviewer'' CodeReviewer~\citep{li2022codereviewer}
.
For AUGER, we find two checkpoints in the paper's repository for best \textit{BLEU} and \textit{Perfect Prediction} metrics, whereas the original paper mentioned and evaluated \textit{ROUGE} and \textit{Perfect Prediction} metrics. 
We have considered both checkpoints in our work.
Additionally, we add the special token \textit{<review\_tag>} to the AUGER tokenizer as required for the AUGER models to work.

Subsequently, we fine-tune these pre-trained transformer models to predict the usefulness of \crcs. 
This fine-tuning process allows us to adapt these powerful models to our specific task, enhancing their ability to classify the usefulness of \crcs\ accurately.

\paragraph{Fine-Tuning} 
To fine-tune the pre-trained models with 768-{4000+} dimensional hidden states, we add two additional linear blocks and drop out with ReLU activations for the intermediate layer.
We explore two approaches: i) freeze all the parameters of the pre-trained models and only update additional layers that are added for our specific task of usefulness classification ii) update all the parameters in the models, allowing them to be further fine-tuned on our dataset.
We employ the \textit{Adam} optimizer to update the model parameters, a widely-used optimization algorithm known for its effectiveness in training deep 
neural network models.

In both approaches, we 
choose hyperparameters such as 
\textit{max input length} $\in \{32, 128\}$  
, \textit{learning rate} $\in \{1e-3, 2e-05\}$ , and \textit{batch size} $\in \{32, 64\}$. 
For training the models, we use the \textit{negative log-likelihood} loss function, where a \textit{softmax} layer is connected to represent the output class, which corresponds to the usefulness classification.
For evaluation (Section~\ref{method_evaluation}), we adopt a 9-fold training and 1-fold testing approach. 
We train the models for a maximum of 10 epochs and select the best model based on the validation loss computed on 10\% of the training data from the 9 folds. 

\paragraph{Commercial Fine-Tuning}
With similar motivation to use commercial embedding for our usefulness classification task, we adhere to OpenAI's recommendations from documentation and select their updated paid GPT-Base model, \textit{babbage-002}. We transform the data as follows (Listing 1). 

\begin{lstlisting}[language=json, caption={Prompt completion format data for fine tuning babbage-002} ]
{ 
    "prompt": "..CR comment..", 
    "completion": "usefulness-label" 
}
\end{lstlisting}

\begin{lstlisting}[language=json, caption={Conversational chat format data for fine tuning GPT-4o }]
 {"messages":
     [
        {"role": "system", 
        "content": "You are classifying your peer's code review feedback as 'useful' or 'not-useful"},
        
        {"role": "user", 
        "content": "..CR comment.." },
        
        {"role": "assistant", 
        "content": "usefulness_label"}
    ]
}
\end{lstlisting}

To check the maximum potential of their GPT models, we also consider their flagship language model,\textbf{ GPT-4o} (“o” for “omni”), the most advanced GPT model. 
We use their latest snapshot of their best model, \textit{gpt-4o-2024-08-06}.
It has 128K context and an October 2023 knowledge cutoff.
It costs US \$3.750 per 1M input tokens, \$15.000 per 1M output tokens, and \$25.000 per 1M training tokens. 
For GPT-4o, we use the prompt shown in Listing 2.

Here, we transform the train, validation, and test files as OpenAI API specifications. 
Next, we fine-tune the model using the already transformed train and validation stratified 10-fold files with auto-hyperparameter configuration. 
We maintain the same evaluation approach as fine-tuning our open LLMs except for the hyperparameter configuration.

\subsection{Evaluating Prediction Models}
\phantomsection \label{method_evaluation}
As our features focus solely on the code review (CR) comments, we establish baseline models that do not utilize any information from developers' experience or review activity. 
These baseline models serve as a benchmark to assess the contribution and effectiveness of the selected features in predicting the usefulness of \crcs\ independently.%

\subsubsection{Within-Project}
The within-project evaluation provides us with reliable and holistic answers to our first two hypotheses testings (\textbf{H1} for proposed features and \textbf{H2} for untapped featureless approaches).

To evaluate the usefulness prediction, we adopt common evaluation metrics used in existing models: \textit{Accuracy}~\citep{bosu2015characteristics,rahman2017predicting,hasan2021usingCRA,pangsakulyanont2014assessing}, \textit{Precision}~\citep{bosu2015characteristics,rahman2017predicting,hasan2021usingCRA,pangsakulyanont2014assessing,meyers2018dataset}, \textit{Recall}~\citep{bosu2015characteristics,rahman2017predicting,hasan2021usingCRA,pangsakulyanont2014assessing,meyers2018dataset}, \textit{F1-score}~\citep{rahman2017predicting,hasan2021usingCRA,pangsakulyanont2014assessing,meyers2018dataset}, and \textit{Area Under Curve (AUC)}~\citep{meyers2018dataset}.
Since different models report their performance using different metrics and treatments, making direct comparisons difficult, we choose models with publicly available computed input features, such as~\citesrf{rahman2017predicting} and \citesrf{meyers2018dataset} for \RH\ and \CC, respectively, as our baselines. 
For setting baseline scores, we use the classifiers as they are in their respective artifacts/papers, and we utilize the available input features.

As \OD\ does not predict usefulness~\citep{turzo2023makes}, we consider \textit{Majority class classification} as a baseline for \OD. 
We then run the 10-fold Stratified Cross-validation on the respective datasets and report the test scores as baseline scores.

For a fair comparison, we maintain the same 10-fold Stratified Cross-validation for our models. 
Finally, we report our test scores for all the existing metrics.
Previous works primarily evaluate their approaches based on \textit{accuracy, AUC, and F$_1$}. %
However, these main evaluation metrics overlook the importance of false-not-useful predictions, which are equally critical.
Both false-useful predictions and false-not-useful predictions can have significant consequences. 
False-useful predictions can negatively impact developers and the overall MCR process, while false-not-useful predictions can lead to reviewers hesitating to rely on usefulness flagging tools/ models.
Additionally, to provide further insights, we introduce two new metrics for this task: the \textit{Matthews Correlation Coefficient (MCC)} and the \textit{Average Precision Score (AP)}. 
These additional metrics offer a more comprehensive evaluation of our models' performance in predicting the usefulness of \crcs.

\subsubsection{Cross-Project}
We evaluate the cross projects through the models' usefulness prediction performance and models' explanation when they are trained on one dataset and tested on other two datasets.  
\paragraph{Cross-generalization:}
While 10-fold cross-validation provides insights into the generalizability of the models, it may not fully account for the variations across different datasets obtained from diverse sources. 
To further assess the generalization capabilities of our models, we conduct experiments where we train the best models on one dataset and test them on others (see Fig.~\ref{fig_approach}). 
Unlike within-project evaluation, we consider a single evaluation metric, \textit{Matthews Correlation Coefficient (MCC)}, which considers all true and false predictions for both positive and negative classes. 
This cross-project evaluation allows us to examine how well the models can generalize across different datasets from various sources. 
This evaluation setup informs us of the generalizability of \crcs' usefulness prediction models; thus, it verifies our third hypothesis introduced in Section~\ref{sec_introduction}, \textbf{H3}, quantitatively. 

\paragraph{Explaining The Best Models:}
\label{sec_meth_xai}
Finally, we explain the best models across datasets using SHAP (SHapley Additive exPlanation)~\citep{shapNIPS2017_7062} values in log-odds space. 
To achieve this, we train a linear explainer model with the hyperparameters and datasets of the best-performing models identified in our experiment for each dataset. 
We then analyze and report the explanations of the predictions on the remaining datasets using this trained linear explainer model. 
This answers our third hypothesis, \textbf{H3}, pseudo-qualitatively.


\section{Results}
\label{sec_results}

This section reports the outcomes of various experiments outlined in Section~\ref{sec_methodology}.

\subsection{Result from Textual Feature-based Approach}
We briefly explore the properties of textual features on the \RH, \CC, and \OD\ datasets. 
After that, we delve into the three feature selection results we obtain. 
Finally, we report the performance results and key findings derived from the analysis of textual features.

\subsubsection{Feature Analysis}
We present the statistical results on features from both the existing literature and the additional features we introduced in our study.

\begin{table}[htbp]
\centering
\caption{Statistical Properties of Features in \RH, \CC, and \OD}
\label{tbl_fts_corr}
\renewcommand{\arraystretch}{1.3}
{ \scriptsize
 Here FID prefixes are \textit{V: voice, C: code, T: text,} J\textit{: jargon }, and $\rho$\textit{: pearson correlation}

}

 \resizebox{\textwidth}{!}{%
\begin{tabular}{|cc|ccc|ccc|ccc|}
\hline
 &  & \multicolumn{3}{|c|}{p-value} & \multicolumn{3}{c|}{effect size(Cohen's D)} & \multicolumn{3}{c|}{$\rho$(fts, usefulness)} \\ \hline
FID & Feature & \RH & \CC & \OD & \RH & \CC & \OD & \RH & \CC & \OD \\ \hline
V29 & tone & - & $\ast$ & $\ast$ & \cellcolor[HTML]{EFF8F4}-0.052 & \cellcolor[HTML]{BEE5D2}0.166 & \cellcolor[HTML]{6AC297}-0.457 & \cellcolor[HTML]{E4ECF9}-0.026 & \cellcolor[HTML]{FF9595}0.068 & \cellcolor[HTML]{3770D4}-0.173 \\
V39 & distress & - & $\ast$ & $\ast$ & \cellcolor[HTML]{FFFFFF}-0.007 & \cellcolor[HTML]{D4EEE1}0.110 & \cellcolor[HTML]{B5E1CB}-0.231 & \cellcolor[HTML]{FFFFFF}-0.003 & \cellcolor[HTML]{FFB7B7}0.045 & \cellcolor[HTML]{9BB7E9}-0.088 \\
V40 & empathy & - & $\ast$ & $\ast$ & \cellcolor[HTML]{FFFFFF}-0.007 & \cellcolor[HTML]{D4EEE1}0.110 & \cellcolor[HTML]{B5E1CB}-0.231 & \cellcolor[HTML]{FFFFFF}-0.003 & \cellcolor[HTML]{FFB7B7}0.045 & \cellcolor[HTML]{9BB7E9}-0.088 \\
V28 & is-toxic & - & - & $\ast$ & \cellcolor[HTML]{EDF7F2}-0.060 & \cellcolor[HTML]{F5FBF8}0.021 & \cellcolor[HTML]{C3E7D5}-0.186 & \cellcolor[HTML]{E0E9F8}-0.029 & \cellcolor[HTML]{FFEDED}0.009 & \cellcolor[HTML]{AFC6ED}-0.071 \\
V38 & gratitude & - & $\ast$ & $\ast$ & \cellcolor[HTML]{FCFEFD}-0.014 & \cellcolor[HTML]{B2E0C9}-0.238 & \cellcolor[HTML]{CFEBDD}-0.152 & \cellcolor[HTML]{FAFBFE}-0.007 & \cellcolor[HTML]{91B0E7}-0.097 & \cellcolor[HTML]{BED0F1}-0.058 \\
V27 & cr-senti & \textbf{$\ast$} & \textbf{$\ast$} & - & \cellcolor[HTML]{D1ECDF}0.118 & \cellcolor[HTML]{B5E1CB}0.192 & \cellcolor[HTML]{E2F3EB}-0.092 & \cellcolor[HTML]{FFA3A3}0.058 & \cellcolor[HTML]{FF8585}0.078 & \cellcolor[HTML]{D9E4F6}-0.035 \\
J47 & density-refact-xerox & - & \textbf{$\ast$} & - & \cellcolor[HTML]{EEF9F4}0.039 & \cellcolor[HTML]{BBE4D0}0.174 & \cellcolor[HTML]{E3F3EB}-0.092 & \cellcolor[HTML]{FFDEDE}0.019 & \cellcolor[HTML]{FF9090}0.071 & \cellcolor[HTML]{D9E4F6}-0.035 \\
V26 & num-propernouns & - & - & - & \cellcolor[HTML]{E3F3EB}-0.090 & \cellcolor[HTML]{F4FBF8}0.022 & \cellcolor[HTML]{EFF8F4}-0.052 & \cellcolor[HTML]{CEDCF4}-0.044 & \cellcolor[HTML]{FFEDED}0.009 & \cellcolor[HTML]{EBF0FA}-0.020 \\
V25 & num-interjections & - & - & - & \cellcolor[HTML]{ECF8F2}0.044 & \cellcolor[HTML]{EAF6F0}-0.069 & \cellcolor[HTML]{F4FAF7}-0.038 & \cellcolor[HTML]{FFDADA}0.022 & \cellcolor[HTML]{E1EAF8}-0.028 & \cellcolor[HTML]{F1F5FC}-0.015 \\
V24 & polarity & - & \textbf{$\ast$} & - & \cellcolor[HTML]{F4FAF7}-0.040 & \cellcolor[HTML]{AEDEC7}-0.250 & \cellcolor[HTML]{F6FBF9}-0.032 & \cellcolor[HTML]{ECF1FA}-0.019 & \cellcolor[HTML]{8BACE6}-0.101 & \cellcolor[HTML]{F4F7FC}-0.012 \\
J46 & density-msoft-u & - & \textbf{$\ast$} & - & \cellcolor[HTML]{E3F4EC}0.068 & \cellcolor[HTML]{6CC499}0.385 & \cellcolor[HTML]{FFFFFF}-0.005 & \cellcolor[HTML]{FFC8C8}0.034 & \cellcolor[HTML]{FF1111}0.155 & \cellcolor[HTML]{FFFDFD}-0.002 \\
C48 & has-out-snippet & - & \textbf{$\ast$} & \textbf{$\ast$} & \cellcolor[HTML]{FAFDFB}-0.020 & \cellcolor[HTML]{FDFEFE}0.000 & \cellcolor[HTML]{FDFEFE}0.000 & \cellcolor[HTML]{F7F9FD}-0.010 & \cellcolor[HTML]{FFFAFA}0.000 & \cellcolor[HTML]{FFFAFA}0.000 \\
V23 & num-exclamation & - & - & - & \cellcolor[HTML]{FCFEFD}-0.013 & \cellcolor[HTML]{F9FCFB}-0.024 & \cellcolor[HTML]{F8FDFA}0.012 & \cellcolor[HTML]{FBFCFE}-0.007 & \cellcolor[HTML]{F7F9FD}-0.010 & \cellcolor[HTML]{FFF3F3}0.005 \\
V22 & is-confirmatory & - & - & - & \cellcolor[HTML]{EEF9F4}0.039 & \cellcolor[HTML]{E9F6F0}-0.071 & \cellcolor[HTML]{F6FCF9}0.018 & \cellcolor[HTML]{FFDEDE}0.019 & \cellcolor[HTML]{E0E9F8}-0.029 & \cellcolor[HTML]{FFF0F0}0.007 \\
V21 & question-ratio & \textbf{$\ast$} & \textbf{$\ast$} & - & \cellcolor[HTML]{C0E5D3}-0.195 & \cellcolor[HTML]{57BB8A}-0.514 & \cellcolor[HTML]{F1FAF5}0.032 & \cellcolor[HTML]{92B1E7}-0.096 & \cellcolor[HTML]{1155CC}-0.205 & \cellcolor[HTML]{FFE8E8}0.012 \\
J45 & density-satd-potdar & - & \textbf{$\ast$} & - & \cellcolor[HTML]{EEF8F3}-0.058 & \cellcolor[HTML]{D9F0E5}0.096 & \cellcolor[HTML]{EDF8F3}0.042 & \cellcolor[HTML]{E1EAF8}-0.028 & \cellcolor[HTML]{FFC0C0}0.039 & \cellcolor[HTML]{FFE2E2}0.016 \\
V37 & implicature & - & \textbf{$\ast$} & - & \cellcolor[HTML]{FAFDFC}0.008 & \cellcolor[HTML]{EEF8F3}-0.056 & \cellcolor[HTML]{E6F5EE}0.060 & \cellcolor[HTML]{FFF5F5}0.004 & \cellcolor[HTML]{E8EEFA}-0.023 & \cellcolor[HTML]{FFD8D8}0.023 \\
J44 & density-msoft-nu & \textbf{$\ast$} & \textbf{$\ast$} & - & \cellcolor[HTML]{DEF1E8}-0.105 & \cellcolor[HTML]{ADDDC5}-0.255 & \cellcolor[HTML]{E6F5EE}0.062 & \cellcolor[HTML]{C6D6F2}-0.051 & \cellcolor[HTML]{89AAE5}-0.103 & \cellcolor[HTML]{FFD7D7}0.024 \\
C20 & code-word-ratio & \textbf{$\ast$} & \textbf{$\ast$} & \textbf{$\ast$} & \cellcolor[HTML]{C6E8D7}0.147 & \cellcolor[HTML]{A5DBC0}0.234 & \cellcolor[HTML]{E5F5ED}0.063 & \cellcolor[HTML]{FF8E8E}0.072 & \cellcolor[HTML]{FF6B6B}0.095 & \cellcolor[HTML]{FFD6D6}0.024 \\
V36 & politeness & - & \textbf{$\ast$} & \textbf{$\ast$} & \cellcolor[HTML]{DBF0E6}-0.115 & \cellcolor[HTML]{C2E6D4}-0.190 & \cellcolor[HTML]{E5F5ED}0.063 & \cellcolor[HTML]{C0D2F1}-0.056 & \cellcolor[HTML]{A7C0EC}-0.077 & \cellcolor[HTML]{FFD6D6}0.024 \\
V19 & num-Qmark & \textbf{$\ast$} & \textbf{$\ast$} & \textbf{$\ast$} & \cellcolor[HTML]{C4E7D6}-0.185 & \cellcolor[HTML]{57BB8A}-0.516 & \cellcolor[HTML]{D9F0E5}0.095 & \cellcolor[HTML]{98B6E9}-0.090 & \cellcolor[HTML]{1155CC}-0.206 & \cellcolor[HTML]{FFC3C3}0.037 \\
V33 & frazier & - & \textbf{$\ast$} & - & \cellcolor[HTML]{E9F6F0}0.054 & \cellcolor[HTML]{F0F9F4}0.035 & \cellcolor[HTML]{D3EEE1}0.111 & \cellcolor[HTML]{FFD3D3}0.026 & \cellcolor[HTML]{FFE5E5}0.014 & \cellcolor[HTML]{FFBBBB}0.042 \\
V18 & rd-text & - & \textbf{$\ast$} & \textbf{$\ast$} & \cellcolor[HTML]{E7F5EE}-0.078 & \cellcolor[HTML]{C2E6D4}-0.191 & \cellcolor[HTML]{C6E8D8}0.145 & \cellcolor[HTML]{D5E1F6}-0.038 & \cellcolor[HTML]{A7C0EC}-0.078 & \cellcolor[HTML]{FFA7A7}0.055 \\
J43 & density-refact-problem & - & - & \textbf{$\ast$} & \cellcolor[HTML]{DAF0E5}0.093 & \cellcolor[HTML]{DDF2E7}0.085 & \cellcolor[HTML]{C2E7D5}0.157 & \cellcolor[HTML]{FFB5B5}0.046 & \cellcolor[HTML]{FFC6C6}0.035 & \cellcolor[HTML]{FFA0A0}0.060 \\
V17 & avg-chars & \textbf{$\ast$} & \textbf{$\ast$} & \textbf{$\ast$} & \cellcolor[HTML]{D4EEE1}0.109 & \cellcolor[HTML]{BFE6D3}0.163 & \cellcolor[HTML]{C1E6D4}0.158 & \cellcolor[HTML]{FFAAAA}0.053 & \cellcolor[HTML]{FF9696}0.067 & \cellcolor[HTML]{FF9F9F}0.061 \\
V32 & pdensity & - & - & \textbf{$\ast$} & \cellcolor[HTML]{F5FBF8}-0.036 & \cellcolor[HTML]{E6F5ED}-0.081 & \cellcolor[HTML]{BDE5D1}0.169 & \cellcolor[HTML]{EDF2FB}-0.018 & \cellcolor[HTML]{DBE5F7}-0.033 & \cellcolor[HTML]{FF9999}0.065 \\
V35 & formality & - & \textbf{$\ast$} & \textbf{$\ast$} & \cellcolor[HTML]{FEFEFE}-0.010 & \cellcolor[HTML]{C4E7D6}0.152 & \cellcolor[HTML]{B8E3CE}0.182 & \cellcolor[HTML]{FDFDFE}-0.005 & \cellcolor[HTML]{FF9E9E}0.062 & \cellcolor[HTML]{FF9292}0.070 \\
V16 & subjectivity & \textbf{$\ast$} & \textbf{$\ast$} & \textbf{$\ast$} & \cellcolor[HTML]{D0ECDE}-0.148 & \cellcolor[HTML]{85CDAA}-0.375 & \cellcolor[HTML]{B8E3CE}0.182 & \cellcolor[HTML]{ADC5ED}-0.072 & \cellcolor[HTML]{5182D9}-0.151 & \cellcolor[HTML]{FF9292}0.070 \\
C15 & has-out-snippet & - & \textbf{$\ast$} & \textbf{$\ast$} & \cellcolor[HTML]{FBFDFC}-0.018 & \cellcolor[HTML]{E4F4EC}-0.087 & \cellcolor[HTML]{B0DFC8}0.203 & \cellcolor[HTML]{F8FAFD}-0.009 & \cellcolor[HTML]{D9E4F6}-0.035 & \cellcolor[HTML]{FF8686}0.078 \\
V31 & cdensity & - & - & \textbf{$\ast$} & \cellcolor[HTML]{FBFEFC}0.006 & \cellcolor[HTML]{EBF7F1}0.047 & \cellcolor[HTML]{AEDFC7}0.209 & \cellcolor[HTML]{FFF6F6}0.003 & \cellcolor[HTML]{FFDDDD}0.019 & \cellcolor[HTML]{FF8282}0.080 \\
J42 & density-secdev & - & \textbf{$\ast$} & \textbf{$\ast$} & \cellcolor[HTML]{FCFDFD}-0.015 & \cellcolor[HTML]{9AD6B8}-0.312 & \cellcolor[HTML]{A1D9BE}0.243 & \cellcolor[HTML]{FAFBFE}-0.007 & \cellcolor[HTML]{6E97E0}-0.126 & \cellcolor[HTML]{FF6F6F}0.093 \\
V14 & avg-punct & \textbf{$\ast$} & - & \textbf{$\ast$} & \cellcolor[HTML]{D0ECDF}0.118 & \cellcolor[HTML]{EEF8F3}0.039 & \cellcolor[HTML]{9ED8BC}0.251 & \cellcolor[HTML]{FFA3A3}0.058 & \cellcolor[HTML]{FFE2E2}0.016 & \cellcolor[HTML]{FF6A6A}0.096 \\
V13 & num-sent & - & \textbf{$\ast$} & \textbf{$\ast$} & \cellcolor[HTML]{F6FCF9}0.019 & \cellcolor[HTML]{68C195}-0.464 & \cellcolor[HTML]{9ED8BB}0.253 & \cellcolor[HTML]{FFECEC}0.009 & \cellcolor[HTML]{2865D1}-0.186 & \cellcolor[HTML]{FF6969}0.097 \\
V12 & stop-word-ratio & - & \textbf{$\ast$} & \textbf{$\ast$} & \cellcolor[HTML]{F9FCFB}-0.024 & \cellcolor[HTML]{B1DFC8}-0.243 & \cellcolor[HTML]{9CD7BA}0.258 & \cellcolor[HTML]{F5F8FC}-0.012 & \cellcolor[HTML]{8FAFE7}-0.098 & \cellcolor[HTML]{FF6666}0.099 \\
V11 & num-tentative & - & \textbf{$\ast$} & \textbf{$\ast$} & \cellcolor[HTML]{E7F5EE}-0.077 & \cellcolor[HTML]{B3E0CA}-0.235 & \cellcolor[HTML]{98D6B7}0.268 & \cellcolor[HTML]{D6E2F6}-0.038 & \cellcolor[HTML]{93B1E7}-0.095 & \cellcolor[HTML]{FF6161}0.102 \\
V34 & informativeness & - & - & \textbf{$\ast$} & \cellcolor[HTML]{F5FBF9}0.020 & \cellcolor[HTML]{E4F4EC}0.066 & \cellcolor[HTML]{94D4B4}0.279 & \cellcolor[HTML]{FFECEC}0.010 & \cellcolor[HTML]{FFD2D2}0.027 & \cellcolor[HTML]{FF5A5A}0.107 \\
C10 & programming-words & \textbf{$\ast$} & - & \textbf{$\ast$} & \cellcolor[HTML]{FAFDFC}-0.020 & \cellcolor[HTML]{DAF0E5}-0.117 & \cellcolor[HTML]{88CFAC}0.312 & \cellcolor[HTML]{F7F9FD}-0.010 & \cellcolor[HTML]{CBD9F3}-0.047 & \cellcolor[HTML]{FF4848}0.119 \\
V9 & num-adverb & - & \textbf{$\ast$} & \textbf{$\ast$} & \cellcolor[HTML]{FDFEFD}-0.012 & \cellcolor[HTML]{81CCA7}-0.388 & \cellcolor[HTML]{85CEAA}0.318 & \cellcolor[HTML]{FCFCFE}-0.006 & \cellcolor[HTML]{4B7ED8}-0.156 & \cellcolor[HTML]{FF4444}0.121 \\
V8 & num-adj & \textbf{$\ast$} & \textbf{$\ast$} & \textbf{$\ast$} & \cellcolor[HTML]{CDEADC}-0.158 & \cellcolor[HTML]{89CFAD}-0.363 & \cellcolor[HTML]{82CDA8}0.327 & \cellcolor[HTML]{A8C1EC}-0.077 & \cellcolor[HTML]{5787DB}-0.146 & \cellcolor[HTML]{FF3F3F}0.125 \\
V30 & yngve & - & \textbf{$\ast$} & \textbf{$\ast$} & \cellcolor[HTML]{E2F3EB}-0.092 & \cellcolor[HTML]{62BF92}-0.481 & \cellcolor[HTML]{7CCAA4}0.344 & \cellcolor[HTML]{CDDBF4}-0.045 & \cellcolor[HTML]{2060CF}-0.193 & \cellcolor[HTML]{FF3636}0.131 \\
V7 & num-nouns & - & \textbf{$\ast$} & \textbf{$\ast$} & \cellcolor[HTML]{E9F6F0}-0.072 & \cellcolor[HTML]{81CCA7}-0.386 & \cellcolor[HTML]{7AC9A2}0.348 & \cellcolor[HTML]{D9E4F6}-0.035 & \cellcolor[HTML]{4C7FD8}-0.156 & \cellcolor[HTML]{FF3333}0.132 \\
V6 & num-determinants & - & \textbf{$\ast$} & \textbf{$\ast$} & \cellcolor[HTML]{E3F3EB}-0.091 & \cellcolor[HTML]{71C59C}-0.434 & \cellcolor[HTML]{70C69C}0.373 & \cellcolor[HTML]{CEDCF4}-0.045 & \cellcolor[HTML]{366FD3}-0.175 & \cellcolor[HTML]{FF2525}0.142 \\
V5 & num-chars & - & \textbf{$\ast$} & \textbf{$\ast$} & \cellcolor[HTML]{F6FBF9}-0.033 & \cellcolor[HTML]{6AC297}-0.458 & \cellcolor[HTML]{69C397}0.391 & \cellcolor[HTML]{F0F4FB}-0.016 & \cellcolor[HTML]{2B67D1}-0.184 & \cellcolor[HTML]{FF1B1B}0.149 \\
V4 & word-count & - & \textbf{$\ast$} & \textbf{$\ast$} & \cellcolor[HTML]{EBF6F1}-0.067 & \cellcolor[HTML]{5ABC8C}-0.505 & \cellcolor[HTML]{69C397}0.392 & \cellcolor[HTML]{DCE6F7}-0.033 & \cellcolor[HTML]{1658CD}-0.202 & \cellcolor[HTML]{FF1A1A}0.149 \\
V3 & avg-stopwords & - & \textbf{$\ast$} & \textbf{$\ast$} & \cellcolor[HTML]{DFF2E9}-0.102 & \cellcolor[HTML]{82CCA8}-0.384 & \cellcolor[HTML]{67C295}0.398 & \cellcolor[HTML]{C7D7F3}-0.050 & \cellcolor[HTML]{4D7FD8}-0.155 & \cellcolor[HTML]{FF1717}0.151 \\
J41 & density-refact-solution & - & \textbf{$\ast$} & \textbf{$\ast$} & \cellcolor[HTML]{FEFEFE}-0.009 & \cellcolor[HTML]{79C8A1}-0.412 & \cellcolor[HTML]{61BF91}0.415 & \cellcolor[HTML]{FDFEFE}-0.004 & \cellcolor[HTML]{4076D6}-0.166 & \cellcolor[HTML]{FF0E0E}0.157 \\
V2 & num-verb & - & \textbf{$\ast$} & \textbf{$\ast$} & \cellcolor[HTML]{F0F9F5}0.033 & \cellcolor[HTML]{85CDAA}-0.375 & \cellcolor[HTML]{61BF91}0.415 & \cellcolor[HTML]{FFE2E2}0.016 & \cellcolor[HTML]{5182D9}-0.151 & \cellcolor[HTML]{FF0E0E}0.157 \\
V1 & avg-words & - & \textbf{$\ast$} & \textbf{$\ast$} & \cellcolor[HTML]{E7F5EE}-0.078 & \cellcolor[HTML]{98D5B7}-0.318 & \cellcolor[HTML]{57BB8A}0.439 & \cellcolor[HTML]{D6E1F6}-0.038 & \cellcolor[HTML]{6B95DF}-0.129 & \cellcolor[HTML]{FF0000}0.166 \\ \hline
\end{tabular}
}
\end{table}

 \begin{table}[htbp]
\caption{Set of Features from our Feature-Selection Approach}
\label{tbl_fts_sel}
\centering
\footnotesize{ Here, 
\textbf{sig}: Statistically significant,  
\textbf{rel}: RFECV relevant, and 
\textbf{imp}: XGB important features. FID prefixes are \textit{V: voice, C: code, T: text,} J\textit{: jargon }

}

 \resizebox{\textwidth}{!}{%
\renewcommand{\arraystretch}{1.3}
\begin{tabular}{|ll|m{0.7cm}m{0.7cm}m{0.7cm}|m{0.7cm}m{0.7cm}m{0.7cm}|m{0.7cm}m{0.7cm}m{0.7cm}|}
\hline
 &  &  \multicolumn{3}{c}{\textbf{Sig}nificant}& \multicolumn{3}{|c}{\textbf{Rel}evant} & \multicolumn{3}{|c|}{\textbf{Imp}ortant} \\ \cline{3-11}
FID & Feature &  \RH & \CC & \OD  & \RH & \CC & \OD & \RH & \CC & \OD \\ \hline 
V1 & avg-words & \FALSE & \TRUE & \TRUE & \FALSE & \FALSE & \TRUE & \FALSE & \FALSE & \FALSE \\
V2 & num-verb & \FALSE & \TRUE & \TRUE & \TRUE & \TRUE & \TRUE & \FALSE & \FALSE & \FALSE \\
V3 & avg-stopwords & \FALSE & \TRUE & \TRUE & \TRUE & \TRUE & \FALSE & \FALSE & \FALSE & \FALSE \\
V4 & word-count & \FALSE & \TRUE & \TRUE & \TRUE & \TRUE & \FALSE & \FALSE & \FALSE & \FALSE \\
V5 & num-chars & \FALSE & \TRUE & \TRUE & \FALSE & \FALSE & \FALSE & \TRUE & \TRUE & \TRUE \\
V6 & num-determinants & \FALSE & \TRUE & \TRUE & \TRUE & \TRUE & \FALSE & \FALSE & \FALSE & \FALSE \\
V7 & num-nouns & \FALSE & \TRUE & \TRUE & \TRUE & \TRUE & \FALSE & \FALSE & \FALSE & \FALSE \\
V8 & num-adj & \TRUE & \TRUE & \TRUE & \TRUE & \TRUE & \TRUE & \FALSE & \FALSE & \FALSE \\
V9 & num-adverb & \FALSE & \TRUE & \TRUE & \TRUE & \TRUE & \TRUE & \FALSE & \FALSE & \FALSE \\
C10 & programming-words & \TRUE & \FALSE & \TRUE & \TRUE & \TRUE & \TRUE & \FALSE & \FALSE & \FALSE \\
V11 & num-tentative & \FALSE & \TRUE & \TRUE & \TRUE & \FALSE & \TRUE & \FALSE & \FALSE & \FALSE \\
V12 & stop-word-ratio & \FALSE & \TRUE & \TRUE & \TRUE & \TRUE & \TRUE & \FALSE & \TRUE & \FALSE \\
V13 & num-sent & \FALSE & \TRUE & \TRUE & \TRUE & \TRUE & \TRUE & \FALSE & \FALSE & \FALSE \\
V14 & avg-punct & \TRUE & \FALSE & \TRUE & \TRUE & \TRUE & \TRUE & \TRUE & \FALSE & \TRUE \\
C15 & has-out-snippet & \FALSE & \TRUE & \TRUE & \TRUE & \FALSE & \TRUE & \FALSE & \FALSE & \FALSE \\
V16 & subjectivity & \TRUE & \TRUE & \TRUE & \TRUE & \TRUE & \TRUE & \FALSE & \FALSE & \FALSE \\
V17 & avg-chars & \TRUE & \TRUE & \TRUE & \TRUE & \TRUE & \TRUE & \TRUE & \TRUE & \TRUE \\
V18 & rd-text & \FALSE & \TRUE & \TRUE & \FALSE & \FALSE & \FALSE & \FALSE & \FALSE & \FALSE \\
V19 & num-Qmark & \TRUE & \TRUE & \TRUE & \TRUE & \TRUE & \TRUE & \FALSE & \FALSE & \FALSE \\
C20 & code-word-ratio & \TRUE & \TRUE & \TRUE & \TRUE & \TRUE & \TRUE & \FALSE & \FALSE & \FALSE \\
V21 & question-ratio & \TRUE & \TRUE & \FALSE & \TRUE & \TRUE & \FALSE & \FALSE & \FALSE & \FALSE \\
V22 & is-confirmatory & \FALSE & \FALSE & \FALSE & \TRUE & \FALSE & \TRUE & \FALSE & \FALSE & \FALSE \\
V23 & num-exclamation & \FALSE & \FALSE & \FALSE & \FALSE & \FALSE & \TRUE & \FALSE & \FALSE & \FALSE \\
V24 & polarity & \FALSE & \TRUE & \FALSE & \TRUE & \TRUE & \TRUE & \FALSE & \FALSE & \FALSE \\
V25 & num-interjections & \FALSE & \FALSE & \FALSE & \FALSE & \FALSE & \TRUE & \FALSE & \FALSE & \FALSE \\
V26 & num-propernouns & \FALSE & \FALSE & \FALSE & \FALSE & \FALSE & \TRUE & \FALSE & \FALSE & \FALSE \\
V27 & cr-senti & \TRUE & \TRUE & \FALSE & \TRUE & \FALSE & \TRUE & \FALSE & \FALSE & \FALSE \\
V28 & is-toxic & \FALSE & \FALSE & \TRUE & \FALSE & \FALSE & \TRUE & \FALSE & \FALSE & \FALSE \\
V29 & tone & \FALSE & \TRUE & \TRUE & \TRUE & \FALSE & \TRUE & \FALSE & \FALSE & \FALSE \\
V30 & yngve & \FALSE & \TRUE & \TRUE & \TRUE & \TRUE & \TRUE & \TRUE & \TRUE & \TRUE \\
V31 & cdensity & \FALSE & \FALSE & \TRUE & \TRUE & \TRUE & \TRUE & \TRUE & \FALSE & \TRUE \\
V32 & pdensity & \FALSE & \FALSE & \TRUE & \TRUE & \TRUE & \TRUE & \FALSE & \TRUE & \FALSE \\
V33 & frazier & \FALSE & \TRUE & \FALSE & \TRUE & \TRUE & \TRUE & \TRUE & \TRUE & \TRUE \\
V34 & informativeness & \FALSE & \FALSE & \TRUE & \TRUE & \TRUE & \TRUE & \TRUE & \TRUE & \TRUE \\
V35 & formality & \FALSE & \TRUE & \TRUE & \TRUE & \TRUE & \TRUE & \TRUE & \TRUE & \TRUE \\
V36 & politeness & \FALSE & \TRUE & \TRUE & \TRUE & \TRUE & \TRUE & \TRUE & \TRUE & \TRUE \\
V37 & implicature & \FALSE & \TRUE & \FALSE & \TRUE & \TRUE & \TRUE & \TRUE & \TRUE & \TRUE \\
V38 & gratitude & \FALSE & \TRUE & \TRUE & \FALSE & \FALSE & \TRUE & \FALSE & \FALSE & \FALSE \\
V39 & distress & \FALSE & \TRUE & \TRUE & \FALSE & \FALSE & \FALSE & \FALSE & \FALSE & \FALSE \\
V40 & empathy & \FALSE & \TRUE & \TRUE & \TRUE & \TRUE & \TRUE & \TRUE & \TRUE & \TRUE \\
J41 & density-refact-solution & \FALSE & \TRUE & \TRUE & \TRUE & \TRUE & \TRUE & \FALSE & \FALSE & \FALSE \\
J42 & density-secdev & \FALSE & \TRUE & \TRUE & \TRUE & \TRUE & \TRUE & \FALSE & \FALSE & \FALSE \\
J43 & density-refact-problem & \FALSE & \FALSE & \TRUE & \TRUE & \TRUE & \TRUE & \FALSE & \FALSE & \FALSE \\
J44 & density-msoft-nu & \TRUE & \TRUE & \FALSE & \TRUE & \FALSE & \TRUE & \FALSE & \FALSE & \FALSE \\
J45 & density-satd-potdar & \FALSE & \TRUE & \FALSE & \TRUE & \FALSE & \TRUE & \FALSE & \FALSE & \FALSE \\
J46 & density-msoft-u & \FALSE & \TRUE & \FALSE & \TRUE & \FALSE & \TRUE & \FALSE & \FALSE & \FALSE \\
J47 & density-refact-xerox & \FALSE & \TRUE & \FALSE & \FALSE & \FALSE & \TRUE & \FALSE & \FALSE & \FALSE \\
C48 & has-out-snippet & \FALSE & \TRUE & \TRUE & \TRUE & \FALSE & \TRUE & \FALSE & \FALSE & \FALSE \\ \hline
\end{tabular}
}
\end{table}

\paragraph{Within-Project Features}
Table~\ref{tbl_fts_corr} presents the input textual features along with their statistical properties related to usefulness. 
We identify significant features (p-value <0.05), including 
\textit{subjectivity, num-adj, code-word-ratio}, and \textit{num-Qmark} in all datasets. 
However, textual features such as \textit{is-confirmatory, num-exclamation, num-interjection}, and \textit{num-propernouns} are not able to distinguish usefulness in any of the datasets. 
These features with p-value <0.05 also serve as one of our three feature selection approaches, namely significant features in Table~\ref{tbl_fts_sel}.

Table~\ref{tbl_fts_sel} shows the list of feature(s) from our three feature-selection approaches. 
Through RFECV Selection~\citep{meyers2018dataset,hasan2021usingCRA}, we select 23 consistently useful features across all datasets. 
Features like \textit{density-refact-xerx, num-propernouns,  num-word/character}, and \textit{distress} are eliminated by all datasets.
Regarding XGB importance-based feature selection, we find that \textit{empathy, implicature, {num-chars}, politeness, informativeness, formality, frazier}, and \textit{yngve} are selected in all datasets, indicating their significance in determining usefulness.

In summary, our analysis reveals that the \textit{avg-chars}, a new syntactic feature, is consistently present in all selection approaches across all datasets. 
Additionally, the proposed features: \textit{avg-chars} and \textit{empathy}, along with existing features: linguistic \textit{politeness, formality}~\citep{meyers2018dataset}, and code-element \textit{stop-word-ratio}~\citep{rahman2017predicting} demonstrate saliency in two or more datasets for each selection approach.

\finding{
Statistical measures and feature selections indicate that the \textit{yngve} readability measure is superior to the commonly used {rd-text},\textit{Flesh-Kincaid} readability formula, for code review \crcs.
All projects consider that objective \crcs\ are more helpful than subjective ones.
}
\paragraph{Cross-Project Features}

Table~\ref{tbl_fts_cross_analysi} shows statistical comparisons among cross-project features' correlation to usefulness experiment described in \ref{fts_analysis_cross_proj}. 
It reveals significant differences in \crcs' feature correlations with usefulness among the Commercial \RH~\citep{rahman2017predicting} and Open Source \CC\ \& \OD~\citep{meyers2018dataset, turzo2023makes} environments. 
\OD\ exhibits significant variations compared to both \RH~($p-value <0.0001,$ $Cohen's$~$D <-1.0$) and \CC~($p-value <0.0001, Cohen's$~$D >1$), while \CC\ and \RH\ show moderate differences between each other ($p-value <0.0001, Cohen's$~$D$\~{}$0.60$).

\begin{table}[ht]
\caption{Comparisons among Projects' Feature Correlations to the Usefulness}
\renewcommand{\arraystretch}{1.3}
\centering
\label{tbl_fts_cross_analysi}
\begin{tabular}{|cc|ccc|}
\hline
\multicolumn{2}{|c|}{\multirow{2}{*}{}} & \multicolumn{3}{c|}{Cohen's D effect-size} \\ \cline{3-5} 
\multicolumn{2}{|c|}{} & \multicolumn{1}{c|}{\CC} & \multicolumn{1}{c|}{\OD} & \RH \\ \hline
\multicolumn{1}{|c|}{\multirow{3}{*}{p-value}} & \CC & \multicolumn{1}{c|}{} & \multicolumn{1}{c|}{-1.278} & -0.637 \\ \cline{2-5} 
\multicolumn{1}{|c|}{} & \OD & \multicolumn{1}{c|}{0.00000783} & \multicolumn{1}{c|}{} & 1.079 \\ \cline{2-5} 
\multicolumn{1}{|c|}{} & \RH & \multicolumn{1}{c|}{0.00012168} & \multicolumn{1}{c|}{0.00000640} &  \\ \hline
\end{tabular}
\end{table}

Upon closer inspection in Table~\ref{tbl_fts_corr}, we observe that \OD\ demonstrates a unique correlation pattern with usefulness. 
It positively correlates social/ interaction-related features and negatively correlates syntactic features. 
In contrast, \CC\ exhibits an opposite behavior, while \RH\ does not display such bipolar characteristics. 
Notably, \OD\ shows a negative correlation of \textit{tone, empathy, distress} with usefulness, whereas \RH\ and \CC\ show less negative and positive correlations for these features, respectively.

A similar antipodal relationship is observed for domain-specific features like {cr-senti} and {is-toxic}. 
When \OD\ and \CC\ are combined, they differ from \RH\ in terms of linguistic features (\textit{informativeness} and \textit{formality}) and the technical feature {density-satd}.

However, all three environments (\OD, \CC, and \RH) exhibit similar associations with features such as \textit{gratitude, polarity,} and {density-refact-prob}. Nevertheless, {density-refact-solu} shows an opposite pattern for \OD\& \CC\ and negligible correlation for \RH.

\finding{%
Social/interaction-related and syntactic features adapted from the general domain exhibit stronger diversity. 
However, similar features from the software domain appear to be weaker and less diverse, particularly in both commercial and open-source environments.
}

\finding{
The usefulness of \crcs\ is negatively correlated with how \crcs\ are presented or expressed in open-source projects, but this correlation is not observed in commercial projects.}

\finding{
Addressing refactoring problems is perceived as somewhat useful in both environments, but the approach to sharing refactoring solutions differs between them.}

\subsubsection{Usefulness Prediction}
We first extract the features and conduct three feature selection procedures: significant (via t-test), relevant (via RFECV), and important (via XGB importance) (Section~\ref{method_evaluation}). 
The sets of features obtained from these procedures are listed in {Table~\ref{tbl_fts_sel}}.

\begin{table*}[htbp]

\caption{{Comparison with Baseline models with \RH}}
\centering
\label{tbl_within_rh}
\begin{sideways}

\begin{tabular}{@{}|llll|rrrrrrrrr|@{}}
\multicolumn{13}{p{1.7\columnwidth}}{
{\scriptsize Here, all \textbf{scores} are reported from models trained on Stratified 10-fold shuffled samples and we kept \textit{Random State=2023} for reproducibility.   }}\\
\multicolumn{13}{p{1.7\columnwidth}}{
{\scriptsize
\textbf{Approaches} \textit{bow:Bag-of-Words, fts:feature-based, emb:featureless-embedding, FT: featureless-fine-tuning an LLM};
\textbf{Config.}
\textit{ft}:\FA, \textit{so}:\SO\ embeddings. Our feature/ feature-sets are described in Table~\ref{tbl_fts_all}\&\ref{tbl_fts_sel}.
\textbf{Metrics }
P: Precision, R: Recall, AUC: Area Under Curve, AP: Average Precision, A: Accuracy, $F_1$-scores, and MCC: Matthews correlation coefficient are reported in differences (\textcolor{darkgray}{-}/\textcolor{black}{+}) with baselines.
}
} \\ 
\hline
 &
   &
   &
   &
  \multicolumn{2}{c}{\textbf{Useful}} &
  \multicolumn{2}{c}{\textbf{Not Useful}} &
  \multicolumn{5}{c|}{\textbf{All}} \\ \cline{5-13}
\textbf{Approach} &
  \textbf{Feature} &
  \textbf{Algorithm/ Model} &
  \textbf{Hyperparameters/ Config} &
  \multicolumn{1}{l}{\textbf{P}} &
  \multicolumn{1}{l}{\textbf{R}} &
  \multicolumn{1}{l}{\textbf{P}} &
  \multicolumn{1}{l}{\textbf{R}} &
  \multicolumn{1}{l}{\textbf{AUC}} &
  \multicolumn{1}{l}{\textbf{AP}} &
  \multicolumn{1}{l}{\textbf{ACC}} &
  \multicolumn{1}{l}{\textbf{F1}} &
  \multicolumn{1}{l|}{\textbf{MCC}} \\ \hline
\hline
\multicolumn{13}{|c|}{\RH$^{}$~\citep{rahman2017predicting}} \\ \hline
Baseline $\rightarrow$ & textual features & \cite{rahman2017predicting}
  & \cite{rahman2017predicting} &
  \textbf{0.616} &
  \textbf{0.766} &
  \textbf{0.464} &
  \textbf{0.297} &
  \textbf{0.543} &
  \textbf{0.629} &
  \textbf{0.576} &
  \textbf{0.681} &
  \textbf{0.071} \\ \hline
  \hline
  closed emb& 
  comment &
  LogisticRegression & 
  text-embedding-3-small &
  \textcolor{black}{\textbf{+.03}} & 
  \textcolor{black}{\textbf{+.08}} & 
  \textcolor{black}{\textbf{+.11}} & 
  \textcolor{black}{\textbf{+.02}} & 
  \textcolor{black}{\textbf{+.11}} & 
  \textcolor{black}{\textbf{+.11}} & 
  \textcolor{black}{\textbf{+.06}} & 
  \textcolor{black}{\textbf{+.05}} & 
  \textcolor{black}{\textbf{+.11}} \\
  
bow &
  text\_tokens &
   LogisticRegression&
  sw-py keywords, stemmed &
  \textcolor{black}{\textbf{+.03}} &
  \textcolor{black}{\textbf{+.09}} &
  \textcolor{black}{\textbf{+.11}} &
  \textcolor{darkgray}{-.01} &
  \textcolor{black}{\textbf{+.08}} &
  \textcolor{black}{\textbf{+.08}} &
  \textcolor{black}{\textbf{+.05}} &
  \textcolor{black}{\textbf{+.05}} &
  \textcolor{black}{\textbf{+.10}} \\
bow &
  text &
   LogisticRegression&
  sw-nonpy keywords, stemmed &
  \textcolor{black}{\textbf{+.03}} &
  \textcolor{black}{\textbf{+.09}} &
  \textcolor{black}{\textbf{+.11}} &
  \textcolor{darkgray}{-.02} &
  \textcolor{black}{\textbf{+.08}} &
  \textcolor{black}{\textbf{+.07}} &
  \textcolor{black}{\textbf{+.05}} &
  \textcolor{black}{\textbf{+.05}} &
  \textcolor{black}{\textbf{+.10}} \\
bow &
  comment &
   LogisticRegression&
   sw-nosw, lemmatized &
  \textcolor{black}{\textbf{+.02}} &
  \textcolor{black}{\textbf{+.13}} &
  \textcolor{black}{\textbf{+.13}} &
  \textcolor{darkgray}{-.07} &
  \textcolor{black}{\textbf{+.08}} &
  \textcolor{black}{\textbf{+.08}} &
  \textcolor{black}{\textbf{+.05}} &
  \textcolor{black}{\textbf{+.06}} &
  \textcolor{black}{\textbf{+.09}} \\
fts &
  rel1 (Table~\ref{tbl_fts_sel})&
   LogisticRegression&
   & 
  \textcolor{black}{\textbf{+.02}} &
  \textcolor{black}{\textbf{+.08}} &
  \textcolor{black}{\textbf{+.08}} &
  \textcolor{darkgray}{-.03} &
  \textcolor{black}{\textbf{+.04}} &
  \textcolor{black}{\textbf{+.03}} &
  \textcolor{black}{\textbf{+.04}} &
  \textcolor{black}{\textbf{+.04}} &
  \textcolor{black}{\textbf{+.07}} \\
open emb &
  ft\_text\_so\_code &
   LogisticRegression&
   \FA, \SO
   &
  \textcolor{black}{\textbf{+.04}} &
  \textcolor{darkgray}{-.07} &
  \textcolor{black}{\textbf{+.03}} &
  \textcolor{black}{\textbf{+.14}} &
  \textcolor{black}{\textbf{+.05}} &
  \textcolor{black}{\textbf{+.06}} &
  \textcolor{black}{\textbf{+.02}} &
  \textcolor{darkgray}{-.01} &
  \textcolor{black}{\textbf{+.07}} \\
open emb &
  so\_text\_ft\_code &
  RandomForest&
  \FA, \SO
   &
  \textcolor{black}{\textbf{+.01}} &
  \textcolor{black}{\textbf{+.12}} &
  \textcolor{black}{\textbf{+.08}} &
  \textcolor{darkgray}{-.10} &
  \textcolor{black}{\textbf{+.04}} &
  \textcolor{black}{\textbf{+.04}} &
  \textcolor{black}{\textbf{+.04}} &
  \textcolor{black}{\textbf{+.05}} &
  \textcolor{black}{\textbf{+.05}} \\
fts &
  all (Table~\ref{tbl_fts_all})&
   LogisticRegression&
   &
  \textcolor{black}{\textbf{+.01}} &
  \textcolor{black}{\textbf{+.10}} &
  \textcolor{black}{\textbf{+.06}} &
  \textcolor{darkgray}{-.07} &
  \textcolor{black}{\textbf{+.04}} &
  \textcolor{black}{\textbf{+.04}} &
  \textcolor{black}{\textbf{+.03}} &
  \textcolor{black}{\textbf{+.04}} &
  \textcolor{black}{\textbf{+.04}} \\
  open-FT &
  comment &
  GraphCodeBERT &
  max-len:32, batch-size:32 ,frozen &
 \textcolor{darkgray}{-.01} &
  \textcolor{black}{\textbf{+.22}} & 
 \textcolor{darkgray}{-.30} &
 \textcolor{darkgray}{-.27} &
  \textcolor{black}{\textbf{+.06}} & 
  \textcolor{black}{\textbf{+.06}} & 
  \textcolor{black}{\textbf{+.03}} & 
  \textcolor{black}{\textbf{+.07}} & 
 \textcolor{darkgray}{-.05}  \\ 
  closed-FT &
  comment &
  GPT-
  babbage-002 & \~{}125B, prompt-completion&
\textcolor{darkgray}{-.01} &
  \textcolor{black}{\textbf{+.18}} & 
  \textcolor{black}{\textbf{+.13}} & 
 \textcolor{darkgray}{-.22} &
 \textcolor{darkgray}{-.} &
 \textcolor{darkgray}{-.} &
  \textcolor{black}{\textbf{+.03}} & 
  \textcolor{black}{\textbf{+.06}} & 
 \textcolor{darkgray}{0} \\ \hline
closed-FT & comment & gpt-4o & \~{}1Tr params, prompt-chat & \textpos{+.06} & \textcolor{darkgray}{-.03} & \textpos{+.07} & \textpos{+.17} & \textcolor{darkgray}{-.} & \textcolor{darkgray}{-.} & \textpos{+.05} & \textpos{+.02} & \textpos{+.13} \\
open-FT & comment & plbart & 410M params, code & \textit{0.00} & \textpos{+.21} & \textpos{+.16} & \textcolor{darkgray}{-.22} & \textpos{+.03} & \textpos{+.03} & \textpos{+.04} & \textpos{+.07} & \textpos{+.04} \\
open-FT & comment & auger & 300M params, CR, PPL & \textit{0.00} & \textpos{+.11} & \textpos{+.02} & \textcolor{darkgray}{-.12} & \textpos{+.02} & \textpos{+.03} & \textpos{+.02} & \textpos{+.04} & \textpos{+.00} \\
open-FT & comment & auger (bleu) & 300M params, CR, BLEU & \textit{0.00} & \textpos{+.13} & \textpos{+.02} & \textcolor{darkgray}{-.15} & \textpos{+.02} & \textpos{+.04} & \textpos{+.02} & \textpos{+.05} & \textcolor{darkgray}{-.02} \\
open-FT & comment & deepseekcoder & 6.7b Instruct, code/CR & \textcolor{darkgray}{-.01} & \textpos{+.23} & \textcolor{darkgray}{-.27} & \textcolor{darkgray}{-.28} & \textcolor{darkgray}{-.01} & \textpos{+.00} & \textpos{+.03} & \textpos{+.07} & \textcolor{darkgray}{-.03} \\
open-FT & comment & codellama & 7B params, code & \textcolor{darkgray}{-.01} & \textpos{+.16} & \textcolor{darkgray}{-.13} & \textcolor{darkgray}{-.22} & \textcolor{darkgray}{-.05} & \textcolor{darkgray}{-.02} & \textpos{+.01} & \textpos{+.05} & \textcolor{darkgray}{-.07} \\
open-FT & comment & codereviewer & 500M params, CR & \textcolor{darkgray}{-.02} & \textpos{+.23} & \textcolor{darkgray}{-.46} & \textcolor{darkgray}{-.30} & \textpos{+.04} & \textpos{+.04} & \textpos{+.02} & \textpos{+.07} & \textcolor{darkgray}{-.07} \\
open-FT & comment & starcoder2 & 7B params, code & \textcolor{darkgray}{-.02} & \textpos{+.23} & \textcolor{darkgray}{-.46} & \textcolor{darkgray}{-.30} & \textcolor{darkgray}{-.03} & \textcolor{darkgray}{-.01} & \textpos{+.02} & \textpos{+.07} & \textcolor{darkgray}{-.07} \\  \hline

\end{tabular}%
\end{sideways}
\end{table*}


\begin{table*}[htbp]

\caption{{Comparison with Baseline models with \CC}}
\centering
\label{tbl_within_cc}
\begin{sideways}

\begin{tabular}{@{}|llll|rrrrrrrrr|@{}}
\multicolumn{13}{p{1.7\columnwidth}}{
{\scriptsize Here, all \textbf{scores} are reported from models trained on Stratified 10-fold shuffled samples and we kept \textit{Random State=2023} for reproducibility.   }}\\
\multicolumn{13}{p{1.7\columnwidth}}{
{\scriptsize
\textbf{Approaches} \textit{bow:Bag-of-Words, fts:feature-based, emb:featureless-embedding, FT: featureless-fine-tuning an LLM};
\textbf{Config.}
\textit{ft}:\FA, \textit{so}:\SO\ embeddings. Our feature/ feature-sets are described in Table~\ref{tbl_fts_all}\&\ref{tbl_fts_sel}.
\textbf{Metrics }
P: Precision, R: Recall, AUC: Area Under Curve, AP: Average Precision, A: Accuracy, $F_1$-scores, and MCC: Matthews correlation coefficient are reported in differences (\textcolor{darkgray}{-}/\textcolor{black}{+}) with baselines.
}
} \\ 
\hline
 &
   &
   &
   &
  \multicolumn{2}{c}{\textbf{Useful}} &
  \multicolumn{2}{c}{\textbf{Not Useful}} &
  \multicolumn{5}{c|}{\textbf{All}} \\ \cline{5-13}
\textbf{Approach} &
  \textbf{Feature} &
  \textbf{Algorithm/ Model} &
  \textbf{Hyperparameters/ Config} &
  \multicolumn{1}{l}{\textbf{P}} &
  \multicolumn{1}{l}{\textbf{R}} &
  \multicolumn{1}{l}{\textbf{P}} &
  \multicolumn{1}{l}{\textbf{R}} &
  \multicolumn{1}{l}{\textbf{AUC}} &
  \multicolumn{1}{l}{\textbf{AP}} &
  \multicolumn{1}{l}{\textbf{ACC}} &
  \multicolumn{1}{l}{\textbf{F1}} &
  \multicolumn{1}{l|}{\textbf{MCC}} \\ \hline
\hline
\multicolumn{13}{|c|}{\CC$^{}$~\cite{meyers2018dataset}} \\ \hline
Baseline $\rightarrow$ 
& all (Table~\ref{tbl_fts_all})& \cite{meyers2018dataset}
  & \cite{meyers2018dataset} &
  \textbf{0.795} &
  \textbf{0.987} &
  \textbf{0.534} &
  \textbf{0.051} &
  \textbf{0.652} &
  \textbf{0.863} &
  \textbf{0.79} &
  \textbf{0.881} &
  \textbf{0.11} \\ \hline
    closed emb &
  comment &
  GausssianNB &
  text-embedding-3-small &
  \textcolor{black}{\textbf{+.11}} & 
 \textcolor{darkgray}{-.30} &
 \textcolor{darkgray}{-.14} &
  \textcolor{black}{\textbf{+.68}} & 
  \textcolor{black}{\textbf{+.12}} & 
  \textcolor{black}{\textbf{+.05}} & 
 \textcolor{darkgray}{-.09} &
 \textcolor{darkgray}{-.10} &
  \textcolor{black}{\textbf{+.24}}\\
fts &
  sig1 (Table~\ref{tbl_fts_sel})&
  GaussianNB &
   &
  \textcolor{black}{\textbf{+.04}} &
  \textcolor{darkgray}{-.15} &
  \textcolor{darkgray}{-.13} &
  \textcolor{black}{\textbf{+.34}} &
  \textcolor{black}{\textbf{+.06}} &
  \textcolor{black}{\textbf{+.02}} &
  \textcolor{darkgray}{-.04} &
  \textcolor{darkgray}{-.04} &
  \textcolor{black}{\textbf{+.13}} \\
bow &
  text\_clean &
   LogisticRegression&
   sw-nonpy keywords, lemmatized &
  \textcolor{black}{\textbf{+.01}} &
  \textcolor{darkgray}{-.01} &
  \textcolor{black}{\textbf{+.13}} &
  \textcolor{black}{\textbf{+.08}} &
  \textcolor{black}{\textbf{+.09}} &
  \textcolor{black}{\textbf{+.04}} &
  \textcolor{black}{\textbf{+.01}} &
  \textcolor{black}{\textbf{+.01}} &
  \textcolor{black}{\textbf{+.12}} \\
fts &
  all &
  RandomForest&
   &
  \textcolor{black}{\textbf{+.01}} &
  \textcolor{darkgray}{0} &
  \textcolor{black}{\textbf{+.15}} &
  \textcolor{black}{\textbf{+.08}} &
  \textcolor{black}{\textbf{+.07}} &
  \textcolor{black}{\textbf{+.03}} &
  \textcolor{black}{\textbf{+.01}} &
  \textcolor{black}{\textbf{+.01}} &
  \textcolor{black}{\textbf{+.12}} \\
bow &
  comment &
   LogisticRegression&
   sw-nonpy keywords, lemmatized &
  \textcolor{black}{\textbf{+.01}} &
  \textcolor{darkgray}{-.01} &
  \textcolor{black}{\textbf{+.09}} &
  \textcolor{black}{\textbf{+.09}} &
  \textcolor{black}{\textbf{+.09}} &
  \textcolor{black}{\textbf{+.04}} &
  \textcolor{black}{\textbf{+.01}} &
  \textcolor{black}{\textbf{+.}} &
  \textcolor{black}{\textbf{+.11}} \\
open emb &
    ft\_text\_ft\_code &

   LogisticRegression&
   \FA & 
  \textcolor{black}{\textbf{+.02}} &
  \textcolor{darkgray}{-.07} &
  \textcolor{darkgray}{-.08} &
  \textcolor{black}{\textbf{+.19}} &
  \textcolor{darkgray}{0} &
  \textcolor{darkgray}{-.02} &
  \textcolor{darkgray}{-.01} &
  \textcolor{darkgray}{-.01} &
  \textcolor{black}{\textbf{+.11}} \\
open emb &
  so\_text\_ft\_code &
   LogisticRegression&
   \FA, \SO
   &
  \textcolor{black}{\textbf{+.02}} &
  \textcolor{darkgray}{-.04} &
  \textcolor{darkgray}{-.06} &
  \textcolor{black}{\textbf{+.14}} &
  \textcolor{black}{\textbf{+.01}} &
  \textcolor{darkgray}{-.01} &
  \textcolor{darkgray}{-.01} &
  \textcolor{darkgray}{-.01} &
  \textcolor{black}{\textbf{+.08}} \\
bow &
  text\_tokens &
   LogisticRegression& 
   sw-py keywords, stemmed &
  \textcolor{darkgray}{-.15} &
  \textcolor{darkgray}{-.13} &
  \textcolor{black}{\textbf{+.04}} &
  \textcolor{black}{\textbf{+.23}} &
  \textcolor{darkgray}{-.03} &
  \textcolor{darkgray}{-.16} &
  \textcolor{darkgray}{-.16} &
  \textcolor{darkgray}{-.15} &
  \textcolor{black}{\textbf{+.06}} \\
bow &
  text &
   LogisticRegression& 
   sw-nonpy keywords, stemmed &
  \textcolor{darkgray}{-.15} &
  \textcolor{darkgray}{-.13} &
  \textcolor{black}{\textbf{+.04}} &
  \textcolor{black}{\textbf{+.23}} &
  \textcolor{darkgray}{-.03} &
  \textcolor{darkgray}{-.16} &
  \textcolor{darkgray}{-.16} &
  \textcolor{darkgray}{-.15} &
  \textcolor{black}{\textbf{+.06}} \\
bow &
  comment &
   LogisticRegression& 
   sw-nosw, lemmatized &
  \textcolor{darkgray}{-.16} &
  \textcolor{darkgray}{-.09} &
  \textcolor{black}{\textbf{+.05}} &
  \textcolor{black}{\textbf{+.18}} &
  \textcolor{darkgray}{-.03} &
  \textcolor{darkgray}{-.16} &
  \textcolor{darkgray}{-.16} &
  \textcolor{darkgray}{-.14} &
  \textcolor{black}{\textbf{+.05}} \\
open-FT &
  comment &
  BERT &
  max-len:32, batch-size:32 ,frozen &
 \textcolor{darkgray}{-.01} &
  \textcolor{black}{\textbf{+.01}} & 
 \textcolor{darkgray}{-.53} &
 \textcolor{darkgray}{-.05} &
  \textcolor{black}{\textbf{+.09}} & 
  \textcolor{black}{\textbf{+.04}} & 
 \textcolor{darkgray}{0} &
 \textcolor{darkgray}{-.0} &
 \textcolor{darkgray}{-.11} \\
  closed-FT &
  comment &
  GPT-
  babbage-002 & \~{}125B, prompt-completion &
  \textcolor{black}{\textbf{+.01}} & 
 \textcolor{darkgray}{-.01} &
  \textcolor{black}{\textbf{+.19}} & 
  \textcolor{black}{\textbf{+.09}} & 
 \textcolor{darkgray}{-.} &
 \textcolor{darkgray}{-.} &
  \textcolor{black}{\textbf{+.01}} & 
 \textcolor{darkgray}{0} &
  \textcolor{black}{\textbf{+.12}}\\
\hline
closed-FT & comment & gpt-4o & \~{}1Tr params, prompt-chat  & \textpos{+.09} & \textcolor{darkgray}{-.04} & \textpos{+.20} & \textpos{+.47} & \textcolor{darkgray}{-.} & \textcolor{darkgray}{-.} & \textpos{+.07} & \textpos{+.03} & \textpos{+.42} \\
open-FT & comment & auger & 300M params, CR, PPL & \textcolor{darkgray}{-.01} & \textpos{+.01} & \textcolor{darkgray}{-.16} & \textcolor{darkgray}{-.04} & \textpos{+.07} & \textpos{+.03} & \textit{0.00} & \textit{0.00} & \textcolor{darkgray}{-.07} \\
open-FT & comment & auger(bleu) & 300M params, CR, BLEU & \textcolor{darkgray}{-.01} & \textpos{+.01} & \textcolor{darkgray}{-.23} & \textcolor{darkgray}{-.05} & \textpos{+.07} & \textpos{+.03} & 0.00 & \textit{0.00} & \textcolor{darkgray}{-.08} \\
open-FT & comment & codellama & 7B params, code & \textcolor{darkgray}{-.01} & \textpos{+.01} & \textcolor{darkgray}{-.43} & \textcolor{darkgray}{-.05} & \textpos{+.01} & \textpos{+.01} & \textit{0.00} & \textit{0.00} & \textcolor{darkgray}{-.10} \\
open-FT & comment & codereviewer & 500M params, CR & \textcolor{darkgray}{-.01} & \textpos{+.01} & \textcolor{darkgray}{-.53} & \textcolor{darkgray}{-.05} & \textpos{+.10} & \textpos{+.05} & 0.00 & \textit{0.00} & \textcolor{darkgray}{-.11} \\
open-FT & comment & deepseekcoder & 6.7b Instruct, code/CR & \textcolor{darkgray}{-.01} & \textpos{+.01} & \textcolor{darkgray}{-.53} & \textcolor{darkgray}{-.05} & \textpos{+.06} & \textpos{+.02} & 0.00 & \textit{0.00} & \textcolor{darkgray}{-.11} \\
open-FT & comment & plbart & 410M params, code & \textcolor{darkgray}{-.01} & \textpos{+.01} & \textcolor{darkgray}{-.53} & \textcolor{darkgray}{-.05} & \textpos{+.07} & \textpos{+.03} & 0.00 & \textit{0.00} & \textcolor{darkgray}{-.11} \\
open-FT & comment & starcoder2 & 7B params,  code & \textcolor{darkgray}{-.01} & \textpos{+.01} & \textcolor{darkgray}{-.53} & \textcolor{darkgray}{-.05} & \textpos{+.04} & \textpos{+.02} & 0.00 & \textit{0.00} & \textcolor{darkgray}{-.11} \\\hline
\end{tabular}%
\end{sideways}
\end{table*}


\begin{table*}[htbp]
\centering
\caption{{Comparison with Baseline models with \OD}}
\label{tbl_within_od}
\begin{sideways}

\begin{tabular}{@{}|llll|rrrrrrrrr|@{}}
\multicolumn{13}{p{1.7\columnwidth}}{
{\scriptsize Here, all \textbf{scores} are reported from models trained on Stratified 10-fold shuffled samples and we kept \textit{Random State=2023} for reproducibility.   }}\\
\multicolumn{13}{p{1.7\columnwidth}}{
{\scriptsize
\textbf{Approaches} \textit{bow:Bag-of-Words, fts:feature-based, emb:featureless-embedding, FT: featureless-fine-tuning an LLM};
\textbf{Config.}
\textit{ft}:\FA, \textit{so}:\SO\ embeddings. Our feature/ feature-sets are described in Table~\ref{tbl_fts_all}\&\ref{tbl_fts_sel}.
\textbf{Metrics }
P: Precision, R: Recall, AUC: Area Under Curve, AP: Average Precision, A: Accuracy, $F_1$-scores, and MCC: Matthews correlation coefficient are reported in differences (\textcolor{darkgray}{-}/\textcolor{black}{+}) with baselines.
}
} \\ 
\hline
 &
   &
   &
   &
  \multicolumn{2}{c}{\textbf{Useful}} &
  \multicolumn{2}{c}{\textbf{Not Useful}} &
  \multicolumn{5}{c|}{\textbf{All}} \\ \cline{5-13}
\textbf{Approach} &
  \textbf{Feature} &
  \textbf{Algorithm/ Model} &
  \textbf{Hyperparameters/ Config} &
  \multicolumn{1}{l}{\textbf{P}} &
  \multicolumn{1}{l}{\textbf{R}} &
  \multicolumn{1}{l}{\textbf{P}} &
  \multicolumn{1}{l}{\textbf{R}} &
  \multicolumn{1}{l}{\textbf{AUC}} &
  \multicolumn{1}{l}{\textbf{AP}} &
  \multicolumn{1}{l}{\textbf{ACC}} &
  \multicolumn{1}{l}{\textbf{F1}} &
  \multicolumn{1}{l|}{\textbf{MCC}} \\ \hline
\hline
\multicolumn{13}{|c|}{\OD$^{}$~\cite{turzo2023makes} } \\ \hline
Baseline $\rightarrow$
& - & \multicolumn{2}{l}{ Majority Class Classification }&

  \textbf{0.82} &
  \textbf{1} &
  \textbf{0} &
  \textbf{0} &
  \textbf{0.50} &
  \textbf{0.82} &
  \textbf{0.82} &
  \textbf{0.9} &
  \textbf{0} \\ \hline
 closed emb & 
 comment & GaussianNB &
 text-embedding-3-small &
  \textcolor{black}{\textbf{+.08}} & 
 \textcolor{darkgray}{-.25} &
  \textcolor{black}{\textbf{+.32}} & 
  \textcolor{black}{\textbf{+.57}} & 
  \textcolor{black}{\textbf{+.19}} & 
  \textcolor{black}{\textbf{+.07}} & 
 \textcolor{darkgray}{-.10} &
 \textcolor{darkgray}{-.09} &
  \textcolor{black}{\textbf{+.26}} \\
bow &
  comment &
  RandomForest&
  sw-scikit, lemmatized &
  \textcolor{black}{\textbf{+.03}} &
  \textcolor{darkgray}{-.02} &
  \textcolor{black}{\textbf{+.55}} &
  \textcolor{black}{\textbf{+.13}} &
  \textcolor{black}{\textbf{+.19}} &
  \textcolor{black}{\textbf{+.09}} &
  \textcolor{black}{\textbf{+.02}} &
  \textcolor{black}{\textbf{+.01}} &
  \textcolor{black}{\textbf{+.21}} \\
bow &
  text\_clean &
  RandomForest&
  sw-scikit, nc &
  \textcolor{black}{\textbf{+.03}} &
  \textcolor{darkgray}{-.02} &
  \textcolor{black}{\textbf{+.57}} &
  \textcolor{black}{\textbf{+.12}} &
  \textcolor{black}{\textbf{+.18}} &
  \textcolor{black}{\textbf{+.08}} &
  \textcolor{black}{\textbf{+.02}} &
  \textcolor{black}{\textbf{+.01}} &
  \textcolor{black}{\textbf{+.21}} \\
bow &
  comment &
  RandomForest&
  sw-nosw, nc &
  \textcolor{black}{\textbf{+.02}} &
  \textcolor{darkgray}{-.02} &
  \textcolor{black}{\textbf{+.64}} &
  \textcolor{black}{\textbf{+.10}} &
  \textcolor{black}{\textbf{+.20}} &
  \textcolor{black}{\textbf{+.09}} &
  \textcolor{black}{\textbf{+.02}} &
  \textcolor{black}{\textbf{+.01}} &
  \textcolor{black}{\textbf{+.20}} \\
fts &
  imp1 (Table~\ref{tbl_fts_sel})&
  GaussianNB &
   &
  \textcolor{black}{\textbf{+.06}} &
  \textcolor{darkgray}{-.35} &
  \textcolor{black}{\textbf{+.25}} &
  \textcolor{black}{\textbf{+.58}} &
  \textcolor{black}{\textbf{+.14}} &
  \textcolor{black}{\textbf{+.07}} &
  \textcolor{darkgray}{-.18} &
  \textcolor{darkgray}{-.15} &
  \textcolor{black}{\textbf{+.18}} \\
bow &
  text\_tokens &
   LogisticRegression& 
   sw-py keywords, stemmed &
  \textcolor{darkgray}{-.18} &
  \textcolor{darkgray}{-.14} &
  \textcolor{black}{\textbf{+.57}} &
  \textcolor{black}{\textbf{+.29}} &
  \textcolor{black}{\textbf{+.12}} &
  \textcolor{darkgray}{-.12} &
  \textcolor{darkgray}{-.19} &
  \textcolor{darkgray}{-.17} &
  \textcolor{black}{\textbf{+.17}} \\
bow &
  text &
   LogisticRegression& 
   sw-nonpy keywords, stemmed &
  \textcolor{darkgray}{-.18} &
  \textcolor{darkgray}{-.14} &
  \textcolor{black}{\textbf{+.57}} &
  \textcolor{black}{\textbf{+.28}} &
  \textcolor{black}{\textbf{+.12}} &
  \textcolor{darkgray}{-.12} &
  \textcolor{darkgray}{-.19} &
  \textcolor{darkgray}{-.17} &
  \textcolor{black}{\textbf{+.17}} \\
bow &
  comment &
   LogisticRegression& 
   sw-nosw, lemmatized &
  \textcolor{darkgray}{-.18} &
  \textcolor{darkgray}{-.11} &
  \textcolor{black}{\textbf{+.59}} &
  \textcolor{black}{\textbf{+.23}} &
  \textcolor{black}{\textbf{+.12}} &
  \textcolor{darkgray}{-.11} &
  \textcolor{darkgray}{-.19} &
  \textcolor{darkgray}{-.16} &
  \textcolor{black}{\textbf{+.16}} \\
open emb &
  so\_text\_so\_code &
   LogisticRegression& 
   \SO
   &
  \textcolor{black}{\textbf{+.03}} &
  \textcolor{darkgray}{-.08} &
  \textcolor{black}{\textbf{+.32}} &
  \textcolor{black}{\textbf{+.18}} &
  \textcolor{black}{\textbf{+.19}} &
  \textcolor{black}{\textbf{+.09}} &
  \textcolor{darkgray}{-.03} &
  \textcolor{darkgray}{-.02} &
  \textcolor{black}{\textbf{+.13}} \\
fts &
  all (Table~\ref{tbl_fts_all})&
  GaussianNB &
   &
  \textcolor{black}{\textbf{+.07}} &
  \textcolor{darkgray}{-.56} &
  \textcolor{black}{\textbf{+.21}} &
  \textcolor{black}{\textbf{+.72}} &
  \textcolor{black}{\textbf{+.15}} &
  \textcolor{black}{\textbf{+.07}} &
  \textcolor{darkgray}{-.33} &
  \textcolor{darkgray}{-.31} &
  \textcolor{black}{\textbf{+.12}} \\
open emb &
  so\_text\_ft\_code &
  GaussianNB &
  \FA, \SO &
  \textcolor{black}{\textbf{+.09}} &
  \textcolor{darkgray}{-.76} &
  \textcolor{black}{\textbf{+.19}} &
  \textcolor{black}{\textbf{+.88}} &
  \textcolor{black}{\textbf{+.17}} &
  \textcolor{black}{\textbf{+.07}} &
  \textcolor{darkgray}{-.47} &
  \textcolor{darkgray}{-.52} &
  \textcolor{black}{\textbf{+.11}} \\
open-FT &
  comment &
  CodeBERT &
  max-len:32, batch-size:32,frozen &
  
  \textcolor{black}{\textbf{+.01}} & 
 \textcolor{darkgray}{0} &
 \textcolor{darkgray}{0} &
 \textcolor{darkgray}{0} &
  \textcolor{black}{\textbf{+.15}} & 
  \textcolor{black}{\textbf{+.07}} & 
  \textcolor{black}{\textbf{+.01}} & 
  \textcolor{black}{\textbf{+.01}} & 
 \textcolor{darkgray}{0}  \\
  closed-FT &
  comment &
  GPT-
  babbage-002 & \~{}125B, prompt-completion &
  \textcolor{black}{\textbf{+.01}} & 
 \textcolor{darkgray}{0} &
 \textcolor{darkgray}{0} &
 \textcolor{darkgray}{0} &
 \textcolor{darkgray}{-.} &
 \textcolor{darkgray}{-.} &
  \textcolor{black}{\textbf{+.01}} & 
  \textcolor{black}{\textbf{+.01}} & 
 \textcolor{darkgray}{0} \\ \hline
 closed-FT & comment & gpt-4o & \~{}1Tr params, prompt-chat & \textpos{+.05} & \textcolor{darkgray}{-.04} & \textpos{+.57} & \textpos{+.26} & \textcolor{darkgray}{-.} & \textcolor{darkgray}{-.} & \textpos{+.02} & \textpos{+.01} & \textpos{+.31} \\
open-FT & comment & auger & 300M params, CR, PPL & \textpos{+.01} & \textcolor{darkgray}{-.01} & \textpos{+.20} & \textpos{+.02} & \textpos{+.16} & \textpos{+.07} & \textpos{+.01} & \textpos{+.01} & \textpos{+.03} \\
open-FT & comment & plbart & 410M params, code & \textpos{+.01} & 0.00 & \textpos{+.10} & \textpos{+.00} & \textpos{+.16} & \textpos{+.08} & \textpos{+.01} & \textpos{+.01} & \textpos{+.02} \\
open-FT & comment & codereviewer & 500M params, CR & \textpos{+.01} & 0.00 & 0.00 & 0.00 & \textpos{+.15} & \textpos{+.06} & \textpos{+.01} & \textpos{+.01} & 0.00 \\
open-FT & comment & deepseekcoder & 6.7b Instruct, code/CR & \textpos{+.01} & 0.00 & 0.00 & 0.00 & \textpos{+.15} & \textpos{+.07} & \textpos{+.01} & \textpos{+.01} & 0.00 \\
open-FT & comment & starcoder2 & 7B params, code & \textpos{+.01} & 0.00 & 0.00 & 0.00 & \textpos{+.08} & \textpos{+.05} & \textpos{+.01} & \textpos{+.01} & 0.00 \\
open-FT & comment & codellama & 7B params, code & \textpos{+.01} & \textcolor{darkgray}{-.01} & \textpos{+.05} & \textpos{+.00} & \textpos{+.14} & \textpos{+.07} & \textpos{+.01} & \textpos{+.01} & \textit{0.00} \\
open-FT & comment & auger(bleu) & 300M params, CR, BLEU & \textpos{+.01} & \textcolor{darkgray}{-.01} & \textpos{+.10} & \textpos{+.00} & \textpos{+.12} & \textpos{+.06} & \textpos{+.01} & \textpos{+.01} & \textcolor{darkgray}{-.01}
\\\hline
\end{tabular}%
\end{sideways}
\end{table*}

%
Next, we train the classifier with these selected feature sets on \RH, \CC, and \OD\ datasets using Stratified 10-fold cross-validation. 
We then evaluate the models' test scores on the stratified 1-fold of respective datasets for the metrics discussed in Section-\ref{method_evaluation}. 
The results in Tables~\ref{tbl_within_rh}, \ref{tbl_within_cc}, and \ref{tbl_within_od} demonstrate that a wide range of textual features and their sets outperform the baselines.

\finding{
Using various feature selection approaches, we achieve higher margins than using all features and outperform all the baselines. 
Interestingly, different feature selection approaches dominate for different datasets.}
\finding{The feature sets supported by at least one dataset emerge in top results, but others do not.}
\summary{
Our \textbf{first hypothesis (H1)} has been confirmed as the approach using textual features outperformed the baseline. 
This observation indicates the textual features' effectiveness in improving performance.
}

\subsection{Result from Featureless Approach} 
In Tables~\ref{tbl_within_rh}, \ref{tbl_within_cc}, and \ref{tbl_within_od}, we present the best results obtained from all of our approaches. 
We describe these results in the following text.

\subsubsection{BoW with TFIDF}
We present the best results from all hyperparameter combinations, including \textit{stopwords, stemming,} and \textit{lemmatization}. 
Tables~\ref{tbl_within_rh}, \ref{tbl_within_cc}, and \ref{tbl_within_od} show that the \BoW(\textit{BoW}) techniques outperform the baseline with the highest or second highest margin for metrics such as \textit{MCC, Accuracy, F$_1$-score,} and \textit{AUC} across all datasets and baselines.
The best hyperparameters for this approach differ for each dataset. 
Notably, the top \textit{MCC} achieving techniques for the three datasets involve three different \textit{stopwords}: \textit{python, non-python,} and \textit{English}. Tables~\ref{tbl_within_rh}, \ref{tbl_within_cc}, and \ref{tbl_within_od} indicates that the selection of \textit{stopwords} {( see Table~\ref{tbl_stopwords})} significantly contributes to improving the scores, along with word-preprocessing methods like \textit{stemming} or \textit{lemmatization}.
In conclusion, the classic BoW-TFIDF technique proves to be powerful for predicting useful \crcs. 
\finding{The importance of \textit{stop words} and word-preprocessing methods varies across different projects.}
\subsubsection{Embedding} 
Tables~\ref{tbl_within_rh}, \ref{tbl_within_cc}, and \ref{tbl_within_od} presents the results for both closed embeddings and open text+code embedding approaches. 
The flagship closed embedding from OpenAI successfully obtains the highest MCC scores for all approaches on all datasets. 
But, they compromise \textbf{9 to 10}\pp\ accuracy and $F_1$ scores, \textbf{25 to 30}\pp\ useful-class-recall scores for \CC\ and \OD. 

With open embeddings, the text+code embedding combinations exhibit non-deterministic results for most of the metrics. 
However, for the text+code combination, \SO+ \SO, \FA+ \FA, and \FA + \SO\ with the \textit{Logistic Regression} classifier achieve the highest \textit{MCC} scores on \OD, \CC, and \RH\ datasets, respectively.

\finding{
For most metrics, closed embedding performs better than open embedding approaches.
Domain-specific open embedding works better than domain-agnostic open embedding for code elements in \crcs.}

\subsubsection{Fine-tuning Transformer Language Models}
Table~\ref{tbl_within_rh},\ref{tbl_within_cc}, and \ref{tbl_within_od} present the best results obtained from fine-tuned advanced GPT-4o, GPT-base, BERT, RoBERTa,
code LLMs: StarCoder-2, PLBart, DeepSeek-Coder, GraphCodeBERT, and CodeBERT models 
code-review specific models: AUGER, CodeReviewer 
for all datasets.
{The latest LLM, GPT-4o has outperformed the rest of the code-specific and code-review-specific LLMs. 
It also performed better than feature-based, embedding-based, and BoW approaches.
Interestingly, GPT-4o has outperformed BoW with only 3\pp\ on commercial \RH, but outperformed on open \CC\ and \OD\ with 20-30\pp\ in terms of MCC metric.
} 
Commercial LLM, GPT-Base has not shown the expected performance for identifying usefulness prediction for \crcs.
The code-specific StarCoder2-7b or DeepSeek-Coder-6.7b and code-review-specific CodeReviewer or AUGER underperform when they are compared to the feature-based or BoW techniques. 
Among all the LLMs, PLBart and AUGER show slightly better performance after GPT-4o. 
The full open language transformer model fine-tuning exhibits catastrophic forgetting, leading to poor performance in classifying the usefulness of \crcs.
However, the performance improves significantly by freezing the transformer network and fine-tuning only the additional layers. 
While these approaches outperform the baseline scores for most evaluation metrics, they do not achieve the highest scores among featureless approaches. 
Notably, they perform well on most evaluation metrics except for \textit{MCC}, which considers not-usefulness.

\finding{
By freezing the transformer parameters, the fine-tuning process avoids catastrophic forgetting.
However, the fine-tuning large language model approach tends to overlook not-useful \crcs, resulting in lower performance in classifying them accurately.
GPT-4o performs better in open-source datasets than in commercial datasets.
}

\finding{
Considering the evaluation metrics, the best featureless approach for predicting the usefulness of \crcs\ is black boxed commercial \textbf{GPT-4o}, then \textbf{BoW} with \textbf{TF-IDF} technique, followed by commercial embedding-based approach, then open text+code embedding-based approach, then the fine-tuning other 
open language models and commercial language models.
}

\summary{
The featureless approach we proposed exhibits superior performance compared to the baseline.
The commercial GPT-4o LLM and non-commercial BoW techniques support our \textbf{second hypothesis (H2)}. 
Furthermore, our flagship \textbf{GPT-4o} based approach 
now serves as the new baseline across all approaches 
and the classical featureless approach (i.e., \textbf{BoW-TFIDF}) serves as the new baseline within non-commercial approaches.

}

\subsection{Result from Within and Cross Projects}
Here, we summarize the within-project results and present the cross-project results. %

\subsubsection{Result from Within-Project Prediction}
\label{result_within}
Our comprehensive evaluation uses a wide range of metrics to determine the best prediction models from each dataset or project.  
Tables~\ref{tbl_within_rh}, \ref{tbl_within_cc}, and \ref{tbl_within_od} show the performance of proposed models from all approaches trained on all deduplicated \crcs\ from the corresponding datasets.
\finding{
The cutting-edge techniques show mixed performance except for flagship GPT-4o. 
After GPT-4o, the naive BoW w/ TF-IDF technique outperforms the baselines with the most metrics without compromising other metrics. 
}
\summary{Overall, the best approach for predicting usefulness of \crcs\ across all datasets, are featuresless \textbf{ GPT-4o }and then \textbf{BoW-TFIDF}. 
}

\subsubsection{Result from Cross-Project Prediction}
{We find the GPT-4o performed the best for most of the metrics in Table~\ref{tbl_within_rh}, \ref{tbl_within_cc}, and ~\ref{tbl_within_od}. 
But Open AI does not share the model thus limiting us to obtain GPT-4o's explanation. 
Therefore, we chose the following best model, the Bag-of-Word technique to evaluate cross-project model performance and model explanations.
}
Then, we train it on all deduplicated \crcs\ from the corresponding datasets using their best hyperparameters found in Tables~\ref{tbl_within_rh}-
\ref{tbl_within_od}.

\paragraph{{Cross-generalization:}}

\begin{table}[ht]
\caption{Comparison on Cross Projects with MCC}

\label{tab:cross_result}
\resizebox{\textwidth}{!}{

\begin{tabular}{@{}ccccccc@{}}
 \hline
\multicolumn{2}{c}{}                                 & \multicolumn{5}{c}{Cross- Project MCC}                               \\ \cline{3-7} 
\multicolumn{2}{c}{\multirow{-2}{*}{Within Project (Best)}} & \RH  & \CC   & \OD  & average        & $\Delta$                      \\ \cline{1-7}
\RH                 & \textbf{0.173}                 &  -    &\textit{ 0.15 } & \textit{0.02 }& \textbf{0.085} & {\color[HTML]{CB0000} -0.088} \\
\CC                 & \textbf{0.233}                 & \textit{0.08} &   -    & \textit{0.01} & \textbf{0.045} & {\color[HTML]{CB0000} -0.188} \\
\OD                 & \textbf{0.207}                 & 0.04 & -0.04 &  -    & \textbf{0}     & {\color[HTML]{CB0000} -0.207} \\ \cline{1-7}

\multicolumn{2}{r}{ \textit{Baseline} $\Rightarrow$} & \textit{0.07} &\textit{ 0.11} & \textit{0}     &    & \\ \cline{3-5}
\end{tabular}
}
\end{table}

Table~\ref{tab:cross_result} compares the \textit{MCC} scores within-projects and the average prediction \textit{MCC} scores on other projects (trained on a full dataset with the best configuration found in Tables~\ref{tbl_within_rh}, \ref{tbl_within_cc}, and \ref{tbl_within_od} and tested on two other full datasets). 
Here, none of the average cross-project performances are higher than the within-project scores ($\Delta < 0$).
Although the model with \RH\ has the lowest within-project \textit{MCC} scores, it shows the least cross-performance loss among other projects.
Specifically, the open-source models seem to be little or not effective for other open-source environments.

However, models from \RH\ \& \CC\ easily outperform the baseline scores for their other projects. 
This observation demonstrates some cross-baseline generalization but not the best cross-project generalization. 
Additionally, the model with \OD\ fails to outperform both baseline scores and within-\OD\ best scores obtained in Table~\ref{tbl_within_od}.

\finding{
In the cross-project evaluation, the models trained on commercial datasets perform slightly better than those trained on open-source datasets, but their generalizability is limited.
}
\summary{
The cross-project evaluation shows limited generalizability. 
Therefore, our \textbf{third hypothesis (H3)} holds for old baselines and commercial projects.

}

\paragraph{{Explanation from the Best Predictors:}}
In accordance with Section~\ref{method_evaluation}, we analyze the explanations of our models using SHAP~\citep{shapNIPS2017_7062} as achieved in Section~\ref{sec_meth_xai}.

\begin{figure}[htbp]
\centerline{\includegraphics[trim={0 0cm 0 0.0cm},clip, width=.9\linewidth]{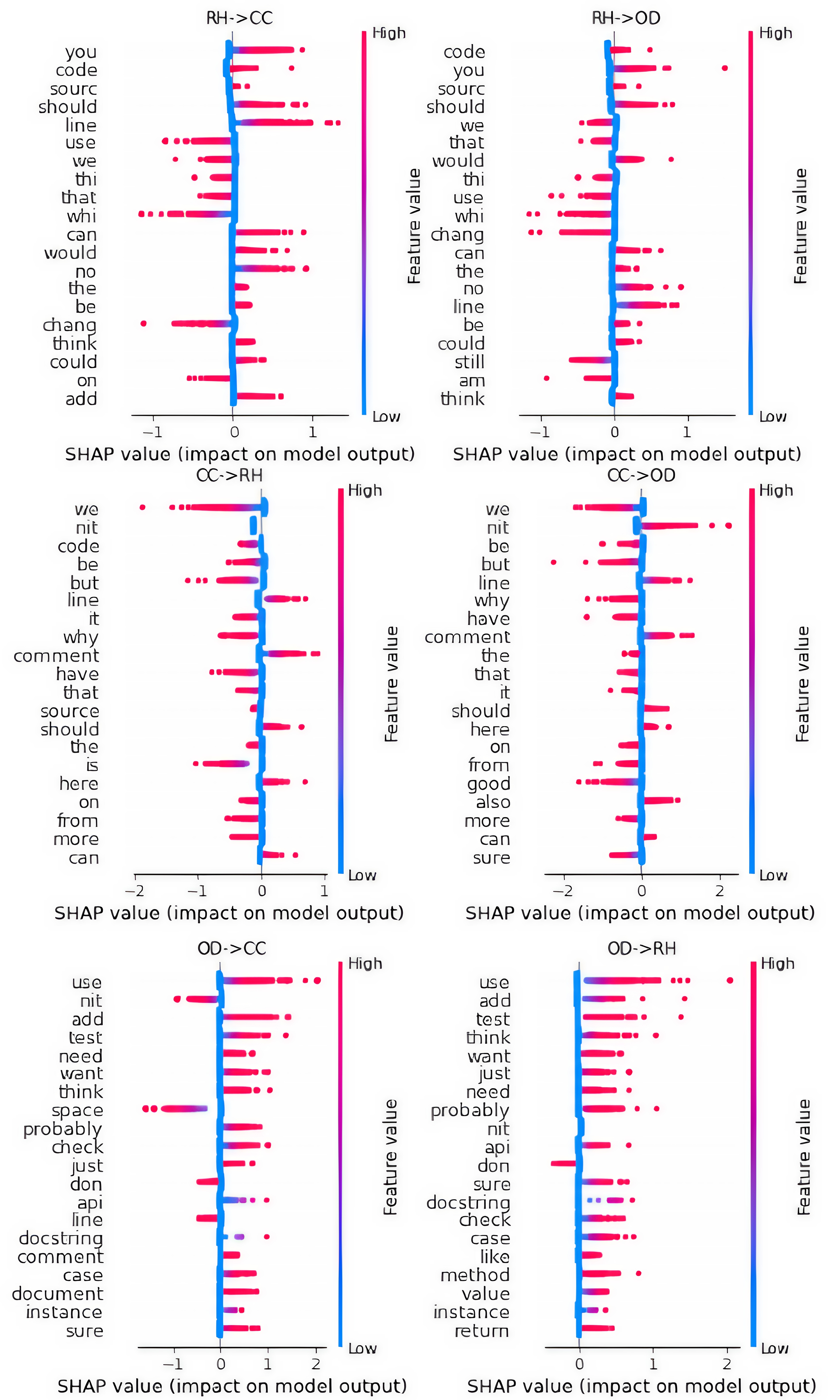}}

\caption{Explaining the Best Models over \RH, \CC, \& \OD} 
\label{fig_shap_vals}
\end{figure}

Fig.~\ref{fig_shap_vals} reveals that words such as \textit{you, should, line,} and \textit{add} have significant contributions to the \RH-model's prediction of \CC-\crcs\ and \OD-\crcs\ as useful.

Intrigued by this finding, we follow excerpting of ~\cite{schneider2016differentiating} and examine \RH\ comments labeled as useful and present two excerpts below.

-\egcrc{This could probably be bumped up one \underline{line} to be with the other properties}

-\egcrc{ Just remove this \underline{line}. There's a specific test case to un-comment, when \\ Zomato starts supporting websites again}

We observe that these words are frequently used to make suggestions and provide guidance on code refactoring.

\finding{
Referring to specific source-code lines or refactoring suggestions are considered useful \crcs.
Also, subjective pronouns `we' and `you' affect the predictions inversely.
}

The \RH-model and \CC-model have common words for both positive and negative SHAP values. 
However, the \CC-model considers words like \textit{line} and \textit{nit} as useful, whereas the \OD-model considers them as not useful.
Interestingly, while predicting \RH, the \OD-model only considers the word \textit{don*} as not useful.

\finding{
The models trained on different open-source datasets have contrasting views on the usefulness of nit-picking comments. One dataset underscores nit-picking as useful, while the other dataset does not.
}

\summary{
The models trained on different datasets from similar environments (i.e., open-source or commercial) do not necessarily produce similar judgments.
}

\section{Implications of Findings}
\label{sec_discussion}

In this section, we summarize and analyze the findings ({\small \emoji{mag}}) obtained from our experiments on predicting the usefulness of \crcs\ in Section~\ref{sec_results}.

\subsection{Textual Features}
 
The proposed features, incorporating both old and new aspects for predicting the usefulness of \crcs, not only lead to improved model performance~\finRef{5} but also effectively capture the variations among existing datasets. 
Through these features and the multiple analyses and selections performed, we observe interesting insights: 
\begin{enumerate}
    \item
The perception of usefulness~\finRef{1} varies based on the features used, shedding light on how different aspects contribute to the perceived usefulness of \crcs.

\item Relationships between language, presentation, and usefulness of \crcs\ are identified in different settings~\finRef{2,3}, showcasing the nuanced nature of this prediction task.

\item The jargon-density features expose variations in how refactoring issues are perceived across projects or environments~\finRef{4,14}, indicating diverse perspectives on suggesting and addressing refactoring.
 
\end{enumerate}
Considering these findings, we anticipate that incorporating more relevant textual features can further enhance our understanding of the binary usefulness of \crcs. 
The widely used Flesh-Kincaid formula from previous works~\citep{bosu2015characteristics,hasan2021usingCRA,rahman2017predicting,turzo2023makes} can not effectively distinguish usefulness in both datasets as the \textit{Yngve} text complexity assessing feature~\citep{meyers2018dataset}~\finRef{1}.
While \citesrf{turzo2023makes} annotate \crc-groups on \OD\ as fine-grained categories of useful \crcs, we discover that five out of six comment groups do not contain any not-useful \crcs. 
Due to the lack of sufficient negative samples, we do not consider \crc-groups as a feature for classifying usefulness.

Considering the insights gained from effective textual features, 
we believe that augmenting the model with features related to \textit{developers' experience} and \textit{code review context} aspects may 
enhance the model's ability to discern the usefulness of \crcs\ more accurately and comprehensively.


\subsection{Featureless Models} 

To the best of our knowledge, no prior works have attempted to predict usefulness using any of our featureless approaches. 
The \textbf{Bag-of-Words TF-IDF (BoW-TFIDF)} method considers only word frequency, while the large language or transformer models and pre-trained embeddings consider word syntax, semantics, and positions. 
Only recently released\textbf{ GPT-4o }with 1 trillion parameters has outperformed as expected. 
Perhaps more exposure to public code and data during GPT-4o' training has empowered \textbf{GPT-4o} to classify well in \CC\&\OD\ from open-source than \RH\ from the commercial environment~\finRef{9}. 
Even code/code-review adapted LLMs ranging from 500 million to \~{}7 billion parameters with frozen parameters have underperformed than the naive BoW-TFIDF approach.  
However, the underperforming models may perform better with larger usefulness labeled datasets, which are unfortunately not available at present. 

Our experiment results partially approves our initial assumption and demonstrates that the less complex featureless approach is more effective for predicting usefulness~\finRef{9,10,12} apart from commercial GPT-4o. 
It also suggests that distinctive words hold more predictive power than complex semantics, irrespective of the domain of the pre-trained models~\finRef{12}. 
This observation is further supported by other findings, such as \textbf{(i)} the impact of word preprocessing techniques like lemmatization, stemming, and stopwords removal on the performance of \textbf{BoW-TFIDF}~\finRef{7}, and \textbf{(ii)} the significance of textual jargon-density features~\finRef{4}.

\subsection{Cross-Project Generalization}
The cross-project evaluation of models using the best hyperparameters indicates that models trained on one project are generally more applicable to other projects, as they mostly outperform the baselines~\finRef{13}. 
However, the average cross-project performance ($\Delta < 0$) suggests that none of the models trained on one project universally apply to all others.

In the feature analysis, we observe a bipolar behavior in \CC\ and \OD~\finRef{2}. The intrinsic inverse nature of the datasets is a prominent reason for the poor performance. 
However, we cannot fully characterize the Chromium and OpenDev communities inversely for two reasons. 
First, the approach used for usefulness annotation differs: semi-automatic for \CC~\citep{meyers2018dataset} and manual for \OD~\citep{turzo2023makes}. 
Second, the latter approach in \OD\ considered both explicit and implicit acceptance or rejection from the developers. 
However, all the annotations for the available datasets \RH, \CC, and \OD\ are not primarily annotated by developers. 
Even primary usefulness annotations from developers may fluctuate from usefulness heuristics from existing research~\citep{bosu2015characteristics,pangsakulyanont2014assessing,turzo2023makes} due to subjectivity issues.
Considering these factors, the performance variations in cross-project evaluation indicate that models' effectiveness for predicting usefulness might vary across projects and communities.

\subsection{Explanation from Models}
The findings from the best performing featureless \textbf{BoW-TFIDF} model~\finRef{14}, which are explained using the SHAP (SHapley Additive exPlanations) framework~\citep{shapNIPS2017_7062}, align well with the results obtained from the analysis of textual features~\finRef{4}. 
This consistency reinforces the significance of our proposed features in predicting usefulness accurately.

However, it is important to note a major limitation of SHAP, which assumes independence of feature values. 
Since \textbf{BoW-TFIDF} also assumes the input vectors to be independent, SHAP's limitation does not hinder the explanation of our best models based on this framework.

\subsection{Datasets} 

In our preliminary analysis, we find that the ratio of \crcs\ to reviewers in the datasets is not reported in the papers~\citep{rahman2017predicting, meyers2018dataset, turzo2023makes}. 
The presence of multiple \crcs\ from a single code reviewer may introduce limitations in non-large datasets.
Upon inspection, we find that only \CC\ has obfuscated 773 email ids, with a maximum of approximately 2\% of the comments coming from a single reviewer. 

To address this issue in future studies, we recommend that researchers consider the ratio of reviewers to \crcs\ and report this information in their papers. 
Moreover, to enhance the robustness of the research findings, future researchers can include more data from various projects and companies, validating and expanding the dataset to benefit the community.

\subsection{Actionable Implications for Developers}
Our findings offer the following actionable implications for developers.

\begin{itemize}

    \item Prioritize composing objective \crcs\ over subjective ones while providing feedback to code authors~\finRef{1}.
    \item Avoid overuse of personal pronouns to maintain a neutral tone~\finRef{13}. 
    \item Craft reviews that are clear, empathetic, and encouraging~\finRef{1, 3, 14}.
    \item Clearly reference specific lines of source-code lines~\finRef{13}.
    \item Provide feedback and share solutions with respect and professionalism~\finRef{1, 3, 4}. 
    \item Use back-ticks to distinguish natural language from programming terms, aiding in \crc\ mining and analysis~\finRef{1, 3, 14}.

\end{itemize}

\section{Threats}
\label{sec_threats}
Here, we address the potential threats to the validity of our study.

\subsection{External Validity} 
The external validity concern in this study is the generalizability of our results. 
To address this threat, we use all three available usefulness annotated datasets, including data from open-source and industry projects. 
We employ stratified cross-validation for within-dataset evaluation to ensure that our models can be generalized effectively.

\subsection{Internal Validity} 
The internal validity pertains to the treatment and outcome.
Existing datasets have availability limitations~\citep{ahmed2023exploring}. 
However, for our experiment design and comparison with baselines, the unavailability of change sets does not affect us as we use textual features from reviewers' \crcs\ for both baselines and proposed models.

It is important to note that the word embeddings used from pre-trained models may carry implicit associations, potentially influencing certain social groups (e.g., race, gender)~\citep{crawford2017trouble,blodgett2020language}. 
As a result, biases could be present in the training and prediction of models or features utilizing these embeddings. 
Our 
Bag-of-Words with TF-IDF approach, being simple, is considered safe from such biases.

\subsection{Construct Validity}
This aspect of validity is focused on theory and observations.
This paper does not provide a specific definition for the usefulness of \crcs. 
Instead, we rely on existing works that have already defined and annotated the usefulness of \crcs.
While the dataset papers have validated their annotation process~\citep{ahmed2023exploring}, there remains a potential threat to the representation of usefulness in the adopted datasets.
Also, we take great care in naming our proposed textual features based on their applicability to ensure accuracy and relevance, thus minimizing this threat.

\subsection{Conclusion Validity}
To draw more accurate and applicable conclusions, we designed the experiment with adequate features and feature-selection approaches.
Next, we applied simple Bag-of-Words to advanced transfer learning from both software and non-software domains.
Then we evaluated the models within and cross-projects trained on deduplicated samples. 
Finally, we verified the best models with a justified, Explainable AI tool. 
Especially, to avoid the threat of conclusion validity, we chose a new metric for this task that considers both false-useful and false-not-useful predictions.
However, our data-driven findings based on available small samples may differ with experiments with large-scale data and language models.

\section{Conclusion}
\label{sec_conclusion}
This paper presents new features and models to predict the usefulness of \crcs\ solely from their content. 
We conduct a thorough analysis of these features and models, examining their generalizability and explanatory power across diverse commercial and open-source datasets.

Through our feature analyses, we gain valuable insights into the perception of binary usefulness.
Our proposed features and feature sets characterize the useful \crc-datasets and describe the underlying usefulness of their \crcs\ that come from different environments and years. 
We observe bipolarity in two of the recent datasets.

Regarding usefulness prediction models, only recent commercial GPT-4o and then
tailored preprocessing-based Bag-of-Word with TF-IDF model outperforms all previous and existing models with the largest margins without code changes or other aspects of features.

Our cross-project experiments show limited generalizability of models trained on the existing datasets. 
Our comprehensive feature analysis and model explanation reveal the bipolarity of datasets, which is evidently restricting the generalizability of models from these datasets. 

The outcomes of this study are expected to significantly contribute to the fields of feature engineering, data mining, and usefulness prediction for \crcs. 
Researchers in the domain of code review will benefit from our work, enabling them to understand and predict the usefulness of \crcs\ better. 
Practitioners can incorporate our data-driven insights in their day-to-day code review practices to improve their process and product.   

In the future, we will expand our promising textual models with code changes, features from developers' experience, and features from code review contexts.
As the large language models require more data samples to perform well, we intend to compile and contribute a large-scale dataset for classifying useful \crcs.
We also plan to conduct an empirical study involving code authors and code reviewers to supplement our findings.

\section{Data Availability}
\label{sec_availability}
The data and artifacts of this paper can be accessed  at -\\
 \textcolor{blue}{\href{https://figshare.com/s/785c17aa02ce22a7db8d}{https://figshare.com/s/785c17aa02ce22a7db8d}}

\section{Conflict of Interest}
The authors declare no conflict of interest.

\bibliographystyle{spbasic}      
\bibliography{references}   

\appendix
\section{Appendix}

\begin{table*}[]
\caption{Programming and non-programming Stop Words}
\label{tbl_stopwords}
\tiny
\begin{longtable}{|p{0.07\linewidth}|p{0.9\linewidth}|}
\hline
\textbf{Adopted From \cite{hasan2021usingCRA}.}

stop word list
& {[}'fifteen',  'own',  'specify',  'inasmuch',  'immediate',  'someday',  'merely',  'behind',  "couldn't",  'did',  'sometime',  'arise',  'recently',  "how's",  'seeming',  'suggest',  'farther',  'yet',  "it'd",  'always',  'different',  'o',  'mostly',  'x',  'opposite',  'etc',  'move',  'whereby',  'successfully',  'okay',  'elsewhere',  "she's",  'nobody',  'two',  "oughtn't",  'which',  'yourselves',  'insofar',  'world',  'previously',  'sub',  'consequently',  'keeps',  'e',  "you'll",  'somebody',  "that's",  "he'll",  'especially',  'looking',  'provides',  'mrs',  'novel',  'secondly',  "that've",  'certainly',  'ex',  'u',  'happens',  'together',  'theyd',  'unless',  'shes',  'arent',  't',  'same',  'miss',  'towards',  'low',  'p',  'owing',  'specified',  'viz',  'taken',  'along',  'everybody',  'the',  'four',  'ca',  'got',  'ord',  'otherwise',  'therere',  'right',  'hundred',  'ltd',  "you'd",  'bottom',  'according',  'is',  'we',  'hence',  'whose',  'section',  'slightly',  'information',  'importance',  'ago',  'thin',  'auth',  'herein',  'amount',  'six',  'appear',  'he',  'much',  's',  'this',  'nine',  'yours',  "won't",  'here',  'those',  'beside',  "i've",  "what've",  'mg',  'necessarily',  'me',  "hadn't",  'call',  'considering',  'minus',  'showns',  "she'll",  "why's",  'hes',  'look',  'are',  'neither',  'fix',  'edu',  "here's",  'howbeit',  "'ll",  'able',  'whither',  'through',  'begin',  'last',  'likewise',  'specifying',  'however',  'youre',  'backwards',  'little',  'wonder',  'normally',  'none',  'page',  'unlike',  'gotten',  'should',  'anyhow',  'know',  'and',  'soon',  'similarly',  'ignored',  'been',  'when',  'inside',  'between',  'aren',  'seeing',  'twenty',  'consider',  'so',  'added',  'hasnt',  'stop',  "didn't",  "mustn't",  'whereafter',  'ref',  'causes',  'widely',  'less',  "we'd",  'neverless',  'besides',  'im',  'getting',  "it's",  'abroad',  'himse”',  'run',  'possible',  'back',  "i'll",  'half',  'end',  'at',  'never',  'sensible',  'throughout',  "a's",  'indicate',  'most',  "where's",  'accordance',  'becomes',  "we've",  'alongside',  'have',  'potentially',  'gives',  'forever',  'please',  'away',  'despite',  'being',  'per',  'make',  'need',  'already',  'mill',  'some',  'new',  'promptly',  'ff',  'c',  'outside',  'obtained',  'inner',  'for',  "shan't",  'entirely',  'kept',  'fify',  'sure',  'home',  'that',  'because',  'id',  'whatever',  'th',  "shouldn't",  'since',  'his',  "who'd",  'regards',  'hereafter',  "ain't",  'showed',  'greetings',  'thence',  'while',  'use',  'whence',  'gone',  're',  'forth',  'let',  'indicated',  'made',  'further',  'better',  'appropriate',  'upon',  'such',  'interest',  'w',  "when's",  'self',  'kg',  'primarily',  'truly',  'once',  "who's",  'latterly',  'placed',  'concerning',  'maybe',  'til',  'ran',  'obtain',  'least',  'everyone',  'rd',  'bill',  'affecting',  'ninety',  'whos',  'also',  'theres',  'don',  'a',  'wish',  'whereas',  'several',  'after',  'above',  'thereof',  "daren't",  'gave',  'million',  'possibly',  "weren't",  'y',  'words',  'whether',  'eight',  'sup',  "'ve",  'empty',  'think',  'de',  'fairly',  'mainly',  'actually',  'resulted',  'three',  'want',  'wherever',  'v',  'what',  'far',  "we're",  'mug',  'largely',  'nevertheless',  'than',  'says',  'thank',  'no-one',  'formerly',  'nowhere',  'no',  'given',  'can',  'detail',  'keys',  'announce',  'like',  'brief',  'come',  'part',  'welcome',  'na',  'by',  'f',  'follows',  'indicates',  'ups',  'ahead',  'appreciate',  'vol',  'regardless',  'another',  'to',  'tell',  'take',  'allows',  'quickly',  'backward',  'looks',  'certain',  'serious',  "doesn't",  'only',  'giving',  'first',  'indeed',  'therefore',  'i',  "there're",  'till',  'refs',  'date',  'about',  'has',  'whod',  'meantime',  'followed',  'seen',  'nor',  'latter',  'course',  "i'd",  'inc',  'various',  "one's",  'system',  'states',  'effect',  'km',  "t's",  'probably',  'perhaps',  'us',  'fill',  'will',  'm',  "she'd",  'toward',  'not',  'itself',  'does',  'changes',  'thou',  'itse”',  'shown',  'ok',  'apart',  'versus',  'herse”',  'begins',  'thirty',  'nos',  'took',  'youd',  'contains',  'might',  "wouldn't",  'saw',  'thats',  'amidst',  'whilst',  'contain',  'except',  'my',  'sincere',  "can't",  'caption',  'comes',  'round',  'dare',  'how',  'noone',  'predominantly',  "hasn't",  'across',  'research',  'against',  'may',  'reasonably',  'anymore',  'described',  "aren't",  'furthermore',  'heres',  'cannot',  'somewhat',  'said',  'forty',  'second',  'seems',  'ever',  'be',  'fifth',  'give',  'biol',  'ts',  "he's",  'invention',  'moreover',  "you're",  'must',  'whim',  'if',  'enough',  'uucp',  'seem',  "they're",  'before',  'z',  'regarding',  'n',  'anyways',  'wed',  'every',  'zero',  'j',  'just',  'downwards',  'needs',  'then',  'say',  'unlikely',  'sometimes',  'went',  'mr',  'particularly',  'later',  'thoroughly',  'makes',  'any',  'again',  "mightn't",  'con',  'alone',  'ed',  'go',  'example',  'put',  'within',  'using',  'accordingly',  'past',  'awfully',  'around',  'it',  'front',  "who'll",  'whoever',  'rather',  'off',  'inward',  'available',  'mine',  'there',  'instead',  'affected',  'usefulness',  'vols',  'doing',  'clearly',  'd',  'hello',  'became',  'index',  'on',  'computer',  'specifically',  'but',  'amongst',  'known',  'theirs',  'qv',  'thanx',  'hi',  'directly',  "needn't",  'briefly',  'hed',  'adj',  'these',  'done',  'eleven',  "they'd",  'beginning',  'unto',  'aside',  'sufficiently',  "it'll",  'amoungst',  'resulting',  'nay',  'from',  'obviously',  'cant',  'thereby',  'onto',  'each',  'hereby',  'would',  'anyone',  'anyway',  'do',  'theyre',  "wasn't",  'hopefully',  "there'll",  'an',  'everything',  'beforehand',  'h',  'significant',  'g',  'neverf',  'act',  'q',  'necessary',  'associated',  'trying',  'very',  'selves',  'things',  'old',  'often',  'anything',  'whomever',  'liked',  'in',  'adopted',  'saying',  'way',  'wheres',  'state',  'keep',  'who',  'all',  'nearly',  'k',  'anywhere',  'him',  'ah',  'relatively',  'whole',  'hither',  'means',  'thing',  'see',  'whichever',  'strongly',  'more',  'amid',  'used',  'useful',  'thereupon',  'going',  'near',  'usually',  'believe',  'undoing',  'under',  'respectively',  "don't",  'pp',  'our',  'ending',  "let's",  'thoughh',  'still',  'allow',  'following',  'hers',  'line',  'substantially',  'thru',  'tried',  'eg',  'due',  'unfortunately',  'whats',  'thereafter',  'currently',  'itd',  'b',  'others',  'her',  'co',  'shall',  'couldnt',  'namely',  'yes',  'lest',  'whom',  'thousand',  'top',  'present',  'or',  'definitely',  'few',  'myself',  'notwithstanding',  'mean',  'former',  'co.',  'et-al',  'affects',  'try',  'ten',  'hid',  'asking',  'thered',  'et',  'too',  'your',  'thick',  'sent',  'although',  'really',  'goes',  'was',  'therein',  'third',  'whenever',  'tip',  'un',  'plus',  'www',  'thereto',  'via',  'else',  'somewhere',  "that'll",  "haven't",  'myse”',  'uses',  'poorly',  'into',  'proud',  'they',  'side',  'knows',  "c'mon",  'com',  'ours',  'among',  'thanks',  "there's",  'quite',  'upwards',  'shed',  'anybody',  'thorough',  'help',  'thus',  'had',  'ourselves',  'exactly',  'somehow',  'important',  'whereupon',  'full',  "he'd",  'seriously',  'yourself',  'its',  'eighty',  'meanwhile',  'results',  'ie',  'even',  'tries',  'apparently',  'related',  'evermore',  'almost',  'sixty',  'well',  'value',  'abst',  'tends',  "you've",  'likely',  'wherein',  'up',  'where',  'beginnings',  'many',  'readily',  'nd',  'get',  'something',  'now',  'find',  "c's",  "what'll",  "what's",  'seven',  'fire',  'name',  "they've",  'somethan',  'provided',  'below',  'why',  'came',  "we'll",  'pages',  'found',  'everywhere',  'one',  "isn't",  'she',  'their',  "there've",  'hereupon',  'willing',  "mayn't",  'am',  'approximately',  'describe',  'r',  'non',  'recent',  'beyond',  'five',  'overall',  'them',  'himself',  'que',  'out',  'lately',  'usefully',  'inc.',  'ones',  'gets',  'sec',  'corresponding',  'best',  'themselves',  'until',  'particular',  'other',  'containing',  'could',  'cry',  'twelve',  'during',  'forward',  "they'll",  'l',  'sorry',  'afterwards',  'both',  'though',  'hardly',  'nothing',  'seemed',  'significantly',  'noted',  'wants',  'over',  "there'd",  'ask',  'fewer',  'herself',  'were',  'cause',  'twice',  'ml',  'lets',  'next',  'taking',  'throug',  'show',  'immediately',  'presumably',  'as',  'someone',  'with',  'without',  'ought',  'similar',  'you',  'of',  'become',  'I',  'oh',  'lower',  'shows',  'underneath',  'having',  "i'm",  'becoming',  'nonetheless',  'vs',  'either',  'down',  'omitted'{]} \\
c reserved & {[}"auto", "else", "long", "switch", "break", "enum", "register", "typedef", "case", "extern", "return", "union", "char", "float", "short", "unsigned", "const", "for", "signed", "void", "continue", "goto", "sizeof", "volatile", "default", "if", "static", "while", "do", "int", "struct", "\_Packed", "double"{]} \\
cpp reserved & {[}"auto", "break", "case", "char", "continue", "default", "do", "double", "else", "entry", "extern", "float", "for", "goto", "if", "int", "long", "register", "return", "short", "sizeof", "static", "struct", "switch", "typedef", "union", "unsigned", "void", "volatile", "while", "asm", "bool", "catch", "class", "const\_cast", "delete", "dynamic\_cast", "explicit", "false", "friend", "inline", "mutable", "namespace", "new", "operator", "private", "public", "protected", "reinterpret\_cast", "static\_cast", "template", "this", "throw", "true", "try", "typeid", "typename", "using", "virtual", "wchar\_t", "and", "and\_eq", "bitand", "bitor", "compl", "not", "not\_eq", "or", "or\_eq", "xor", "xor\_eq", "cin", "cout", "endl", "include", "INT\_MIN", "INT\_MAX", "iomanip", "iostream", "main", "MAX\_RAND", "npos", "NULL", "std", "string"{]} \\
java reserved & {[}"abstract", "assert", "boolean", "break", "byte", "case", "catch", "char", "class", "const", "continue", "default", "do", "double", "else", "enum", "extends", "final", "finally", "float", "for", "goto", "if", "implements", "import", "instanceof", "int", "interface", "long", "native", "new", "package", "private", "protected", "public", "return", "short", "static", "strictfp", "super", "switch", "synchronized", "this", "throw", "throws", "transient", "try", "void", "volatile", "while"{]} \\ \hline

\end{longtable}
\end{table*}

\begin{table*}[]
\begin{longtable}{|p{0.07\linewidth}|p{0.9\linewidth}|}
\hline    
python reserved & {[}"and", "exec", "not", "assert", "finally", "or", "break", "for", "pass", "class", "from", "print", "continue", "global", "raise", "def", "if", "return", "del", "import", "try", "elif", "in", "while", "else", "is", "with", "except", "lambda", "yield"{]} \\
swift reserved & {[}"associatedtype", "class", "deinit", "enum", "extension", "fileprivate", "func", "import", "init", "inout", "internal", "let", "open", "operator", "private", "protocol", "public", "rethrows", "static", "struct", "subscript", "typealias", "var", "break", "case", "continue", "default", "defer", "do", "else", "fallthrough", "for", "guard", "if", "in", "repeat", "return", "switch", "where", "while", "as", "Any", "catch", "false", "is", "nil", "super", "self", "Self", "throw", "throws", "true", "try", "\#available", "\#colorLiteral", "\#column", "\#else", "\#elseif", "\#endif", "\#error", "\#file", "\#fileID", "\#fileLiteral", "\#filePath", "\#function", "\#if", "\#imageLiteral", "\#line", "\#selector", "\#sourceLocation", "\#warning", "associativity", "convenience", "dynamic", "didSet", "final", "get", "infix", "indirect", "lazy", "left", "mutating", "none", "nonmutating", "optional", "override", "postfix", "precedence", "prefix", "Protocol", "required", "right", "set", "Type", "unowned", "weak", "willSet"{]} \\
others & {[}"for(", "for (", "if(", "if (", "while(", "while (", "switch(", "switch (", "else if", "else\{", "else \{", ".c", ".h", ".java", ".py", ".php", ".html", ".css", ".xml", ".tcp", "sqlite3.o", "sqlite.o", "sqlite2.o", ".sh", ".dat", ".josn", "http", "post" "sconscript", "path", "gcov", "add", "null", "flag", "bug", "newline", "memcpy", "boolean", "int", "integer", "float", "double", "char", "void", "class", "struct", "function", "method", "-\textgreater{}", "pointer", "return", "stack", "queue", "vector", "array", "variable", "\#include", "enum", "\#define", "std", "macro", "header", "api", "risk", "cerr", "namespace", "parameter", "append", "prepend", "static" "indent" "loop" "size\_t" "\%d", "\%s", "\%f", "\%lf", "\%zu", "\textbackslash{}\textbackslash{}n", "print", "cout", "stdlib", "apk", "malloc", "switch", "self.", " \\
& \textbf{all stop} = c reserved+cpp reserved+java reserved+swift reserved+others+stop word list \\
& \textbf{programming keywords} = c reserved+cpp reserved+java reserved+swift reserved+others - stop word list \\ \hline

Scikit & {[}'what', 'its', 'were', 'as', 'serious', 'many', 'i', 'next', 'un', 'through', 'top', 'noone', 'being', 'neither', 'see', 'where', 'around', 'each', 'mill', 'empty', 'keep', 'out', 'more', 'my', 'cry', 'eleven', 'the', 'twelve', 'please', 'ltd', 'about', 'still', 'inc', 'therefore', 'everyone', 'over', 'third', 'already', 'among', 'these', 'might', 'via', 'into', 'two', 'whether', 'either', 'others', 'since', 'all', 'made', 'front', 'such', 'it', 'thin', 'everything', 'must', 'same', 'several', 'whom', 'some', 'found', 'whose', 'most', 'every', 'co', 'thence', 'was', 'whatever', 'once', 'everywhere', 'he', 'namely', 'detail', 'they', 'with', 'within', 'whereas', 'per', 'while', 'sometime', 'her', 'latterly', 'rather', 'beyond', 'put', 'so', 'latter', 'whole', 'name', 'became', 'thick', 'myself', 'otherwise', 'least', 'hundred', 'between', 'throughout', 'even', 'could', 'themselves', 'than', 'after', 'enough', 'an', 'fire', 'however', 'she', 'thereupon', 'a', 'are', 'why', 'first', 'herself', 'only', 'now', 'get', 'then', 'towards', 'elsewhere', 'three', 'eight', 'below', 'somehow', 'alone', 'nothing', 'something', 'thereafter', 'last', 'upon', 'who', 'would', 'another', 'due', 'anything', 'except', 'very', 'whereafter', 'less', 'part', 'fifteen', 'nobody', 'perhaps', 'hence', 'sometimes', 'fill', 'can', 'one', 'system', 'whereupon', 'if', 'no', 'none', 'nowhere', 're', 'wherein', 'interest', 'seemed', 'up', 'four', 'their', 'anyone', 'describe', 'thereby', 'should', 'when', 'cannot', 'six', 'sixty', 'hers', 'am', 'we', 'anyhow', 'toward', 'five', 'therein', 'had', 'seems', 'eg', 'never', 'yours', 'during', 'whence', 'go', 'side', 'whoever', 'nine', 'those', 'own', 'which', 'always', 'ever', 'twenty', 'of', 'how', 'whenever', 'in', 'by', 'onto', 'or', 'ours', 'behind', 'there', 'though', 'cant', 'on', 'someone', 'ten', 'becomes', 'other', 'fifty', 'mostly', 'nor', 'against', 'else', 'without', 'been', 'yourself', 'to', 'seeming', 'anyway', 'before', 'your', 'here', 'amount', 'yourselves', 'along', 'take', 'herein', 'be', 'that', 'and', 'from', 'our', 'amoungst', 'at', 'afterwards', 'have', 'them', 'show', 'amongst', 'this', 'us', 'hereupon', 'itself', 'couldnt', 'become', 'moreover', 'often', 'thus', 'anywhere', 'may', 'mine', 'down', 'forty', 'hereafter', 'yet', 'full', 'further', 'somewhere', 'bottom', 'also', 'done', 'ie', 'is', 'meanwhile', 'under', 'himself', 'move', 'until', 'whither', 'will', 'sincere', 'few', 'off', 'formerly', 'his', 'you', 'above', 'hasnt', 'becoming', 'for', 'almost', 'besides', 'but', 'although', 'con', 'beside', 'give', 'etc', 'me', 'beforehand', 'find', 'seem', 'do', 'both', 'him', 'call', 'whereby', 'again', 'de', 'any', 'wherever', 'bill', 'back', 'because', 'has', 'not', 'former', 'ourselves', 'across', 'well', 'much', 'together', 'thru', 'too', 'hereby', 'nevertheless', 'indeed'{]} \\ \hline

\end{longtable}
\end{table*}

\begin{table*}[]
\begin{longtable}{|p{0.07\linewidth}|p{0.9\linewidth}|}
\hline   
NLTK & {[}'theirs', 'should', 'what', 'its', 'were', 'as', 'when', "shouldn't", 'wouldn', 'hers', 'am', 'we', 'm', 'i', 'had', 'through', 'yours', "should've", 'being', 'during', "wouldn't", 'where', 'each', 'those', 'out', 'd', 'more', 'mustn', 'own', 'which', 'my', 've', 'the', 'of', 'how', 'in', 'about', 'by', 'ours', 'or', 'there', 'on', 'other', 'over', 'hasn', 'nor', 'against', "don't", "that'll", 'these', 't', 'into', 'ain', "needn't", 'couldn', 'shouldn', 'been', 'yourself', 'to', 'don', 'hadn', 'ma', 'all', 'your', 'before', 'it', 'such', 'here', 'shan', 'yourselves', 'doesn', 'that', 'be', 'same', 'and', 'whom', 'from', 'some', 'our', 'most', "haven't", "shan't", 'at', 'was', 'having', "hasn't", 'once', 'he', 'have', 'they', 'them', 's', "doesn't", 'with', 'this', 'while', 'her', 'itself', "won't", 'll', 'so', "mustn't", "you'd", 'myself', 'down', "you'll", 'between', 'further', "didn't", 'is', "aren't", "isn't", 'under', 'themselves', 'just', 'than', 'himself', 'after', 'until', "you've", 'haven', 'will', 'an', 'few', 'off', "it's", 'she', 'did', "wasn't", 'y', 'his', 'are', 'you', 'a', 'aren', 'why', 'above', 'didn', 'won', 'weren', 'for', "hadn't", 'herself', 'only', 'then', 'now', 'does', 'doing', 'but', 'below', 'me', 'isn', 'who', "you're", 'do', 'both', 'very', 'him', 'o', "she's", 'needn', 'again', 'any', 'can', 'not', 'has', 'because', "couldn't", 'wasn', 'if', 'no', "mightn't", 're', 'up', 'ourselves', "weren't", 'their', 'too', 'mightn'{]}
\\ \hline
\end{longtable}
\end{table*}

\clearpage
\section*{Biography and Photo}


\noindent
\begin{minipage}{0.3\textwidth}
       \includegraphics[width=\textwidth]{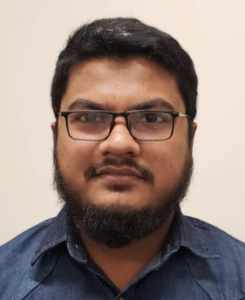} 
\end{minipage}
\hfill
\begin{minipage}{0.65\textwidth}
         \textbf{Sharif Ahmed} is a Computing Ph.D. candidate at Boise State University, Idaho, USA.  His current research interests include Modern Code Review, Software Quality, Software Sustainability, and Software Engineering leveraging Natural Language Processing, Machine Learning, and Deep Learning. Contact him at sharifahmed@u.boisestate.edu
\end{minipage}
\vspace{1em}

\noindent
\begin{minipage}{0.3\textwidth}
    \includegraphics[width=\textwidth]{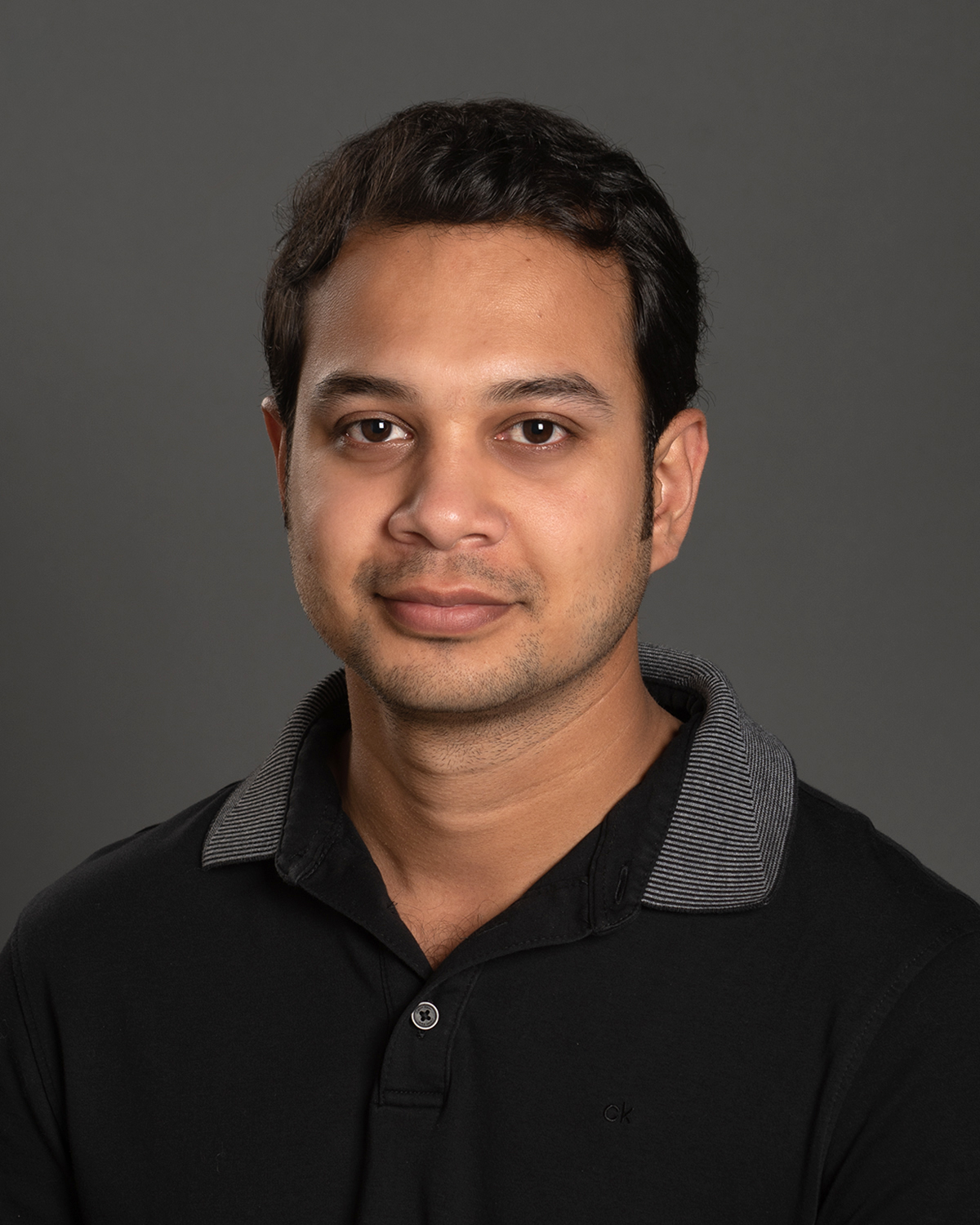} 
\end{minipage}
\hfill
\begin{minipage}{0.65\textwidth}
         \textbf{Nasir U. Eisty} is an Assistant Professor in the Computer Science Department of Boise State University. His research interests lie in the areas of Empirical Software Engineering, AI for Software Engineering, Scientific and Research Software Engineering, and Software Security. He received his Ph.D. degree in Computer Science from the University of Alabama. Contact him at nasireisty@boisestate.edu
\end{minipage}
\end{document}